\documentclass{PoS}

\usepackage{subfig}
\usepackage[sort&compress,numbers,square]{natbib}

\title{Resonances of the systems $\pi^-\eta$ and $\pi^-\eta'$ in the
  reactions $\pi^-p\to\pi^-\eta p_{\textrm{slow}}$ and $\pi^-p\to\pi^-\eta' p_{\textrm{slow}}$ at COMPASS}

\ShortTitle{Resonances of the systems $\pi^-\eta^{(\prime)}$ in the
  reactions $\pi^-p\to\pi^-\eta^{(\prime)} p_{\textrm{slow}}$ at COMPASS}

\author{\speaker{Tobias Schl\"uter}$^{a,c}$, Dmitri Ryabchikov$^b$, Wolfgang D\"unnweber$^a$ and Martin Faessler$^a$%
         \thanks{The authors acknowledge financial support by the German
           Bundesministerium
  f\"ur Bildung und Forschung (BMBF), by the
  Maier-Leibnitz-Laboratorium der LMU und TU M\"unchen, and by the DFG
  cluster of excellence ``Origin and Structure of the Universe.''}\ \   for
the COMPASS collaboration\\
\llap{$^a$}        Ludwig-Maximilians-Universit\"at M\"unchen\\
\llap{$^b$}        Technische Universit\"at M\"unchen and IHEP Protvino\\
\llap{$^c$}        E-mail: \email{tobias.schlueter@physik.uni-muenchen.de}\\
}


     \abstract{We describe partial-wave analyses of the systems
       $\pi^-\eta$ and $\pi^-\eta'$ produced in interactions of a
       $\pi^-$ beam ($190\,\textrm{GeV}/c$) with a liquid hydrogen
       target.  The data were recorded during the 2008 COMPASS run,
       where a slow recoiling proton ($|t|>0.1\,\textrm{GeV}^2$) was
       required by the trigger.  We compare analyses of the
       $\pi^-\eta$ and $\pi^-\eta'$ data.  Significant contributions
       can be attributed to the resonances $a_2(1320)$, observed in
       the $D_+$-wave, and $a_4(2040)$, observed in the $G_+$-wave.
       Between the two systems, we find similar compositions of the
       even partial waves $D_+$ and $G_+$ after taking phase-space
       factors into account, but a much enhanced $P_+$-wave in
       $\pi^-\eta'$.  Relative phase-differences indicate a large
       incoherent contribution of in the $P_+$-wave of the
       $\eta'\pi^-$ system, but other interpretations are not
       excluded.  The known resonances $a_2(1320)$, $a_4(2040)$ and
       their parameters could be extracted from the data; their
       branchings are found to roughly agree with predictions from
       $\eta-\eta'$ mixing.}

\FullConference{Sixth International Conference on Quarks and Nuclear Physics\\
		 April 16-20, 2012\\
		 Ecole Polytechnique, Palaiseau,  Paris}

\begin{document}

\section{Introduction}

Exotic quantum number mesons which cannot be accomodated by $q\overline{q}$
states have been a long sought-for prediction of QCD.  Recent reviews
of the field, which also give references, are
Refs.~\cite{Klempt:2007cp,Meyer:2010ku}.  The
PDG~\cite{Nakamura:2010zzi} lists a spin-exotic $\pi(1400)$ decaying
to $\eta\pi$, and a spin-exotic $\pi(1600)$ decaying to $\eta'\pi$
(both in $P$-wave, with quantum number $J^{PC}=1^{-+}$).  These claims
came surprising not only because of the unexpectedly low mass of the
$\eta\pi$ resonance, but also because hybrid mesons are expected to
preferentially decay into final-states involving $P$-wave mesons such
as $b_1\pi$ or $f_1\pi$, and because by $SU(3)$ arguments a hybrid
meson should prefer decays to $\eta'\pi$ over the $\eta\pi$ channel,
but it should decay to both.  Furthermore the analyses leading to the
PDG entries have been questioned, and alternative theoretical models
have been proposed.

The COMPASS collaboration has extracted large data sets, covering an
unprecendented range of invariant masses, and hopes to clarify the
situation.  In 2008 the experiment~\cite{Abbon:2007pq} took data with
a $190\,\textrm{GeV}$ pion beam impinging on a liquid hydrogen target,
aiming at collecting large samples of data for spectroscopy.  First
results for the $\eta'\pi^-$ system were given at a previous
conference~\cite{Schluter:2011b}.  The data selection is also
described in the reference, up to minor refinements having taken place
in the meantime.  The reactions under consideration are
$\pi^-p\to\pi^-\eta^{(\prime)}p$.  We will focus on the $\eta\pi^-$
system and on the comparison between the two systems.  Additionally,
we will briefly discuss fits to the partial-wave results with
resonance models.  The data for both final states were analyzed with
the same partial-wave software, where the full four-body dynamics of
the $\pi^-\pi^-\pi^+\pi^0$ and $\pi^-\pi^-\pi^+\eta$ systems was taken
into account in order to separate the three-body decays of the
isoscalars from the inevitable background.  Additionally, the data
were analyzed with a two-body program that was also used in another
analysis presented at this conference~\cite{Austregesilo:2012}.  The
results were found to be compatible between the two approaches.

\section{Partial-wave Analysis in Mass Bins}

\begin{figure}[htbp]
  \centering
  
  \subfloat[$|P_+|^2$]{\includegraphics[width=.24\textwidth]{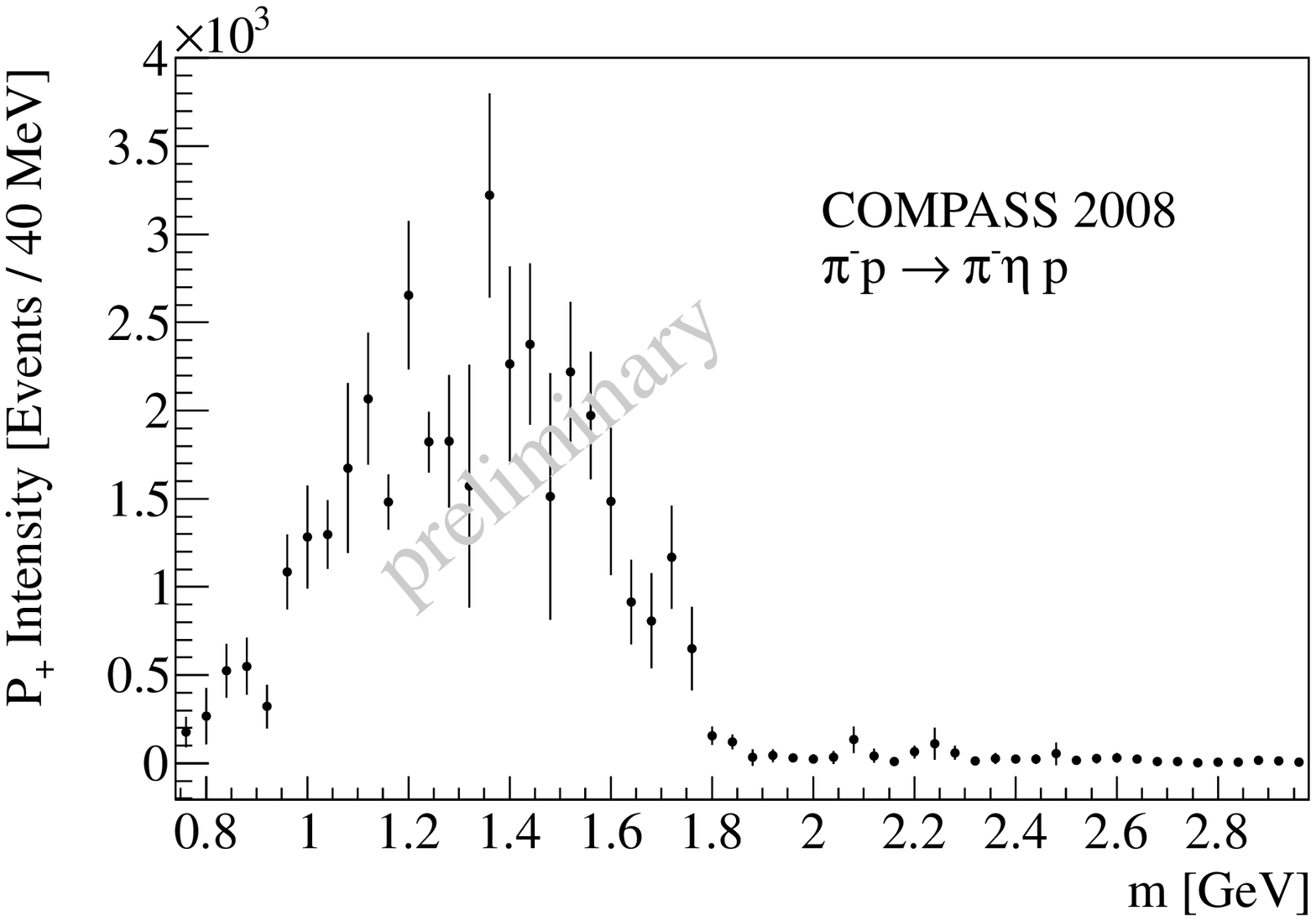}}
  \subfloat[$\textrm{Re}(P_{+}^*D_{+})$]{\includegraphics[width=.24\textwidth]{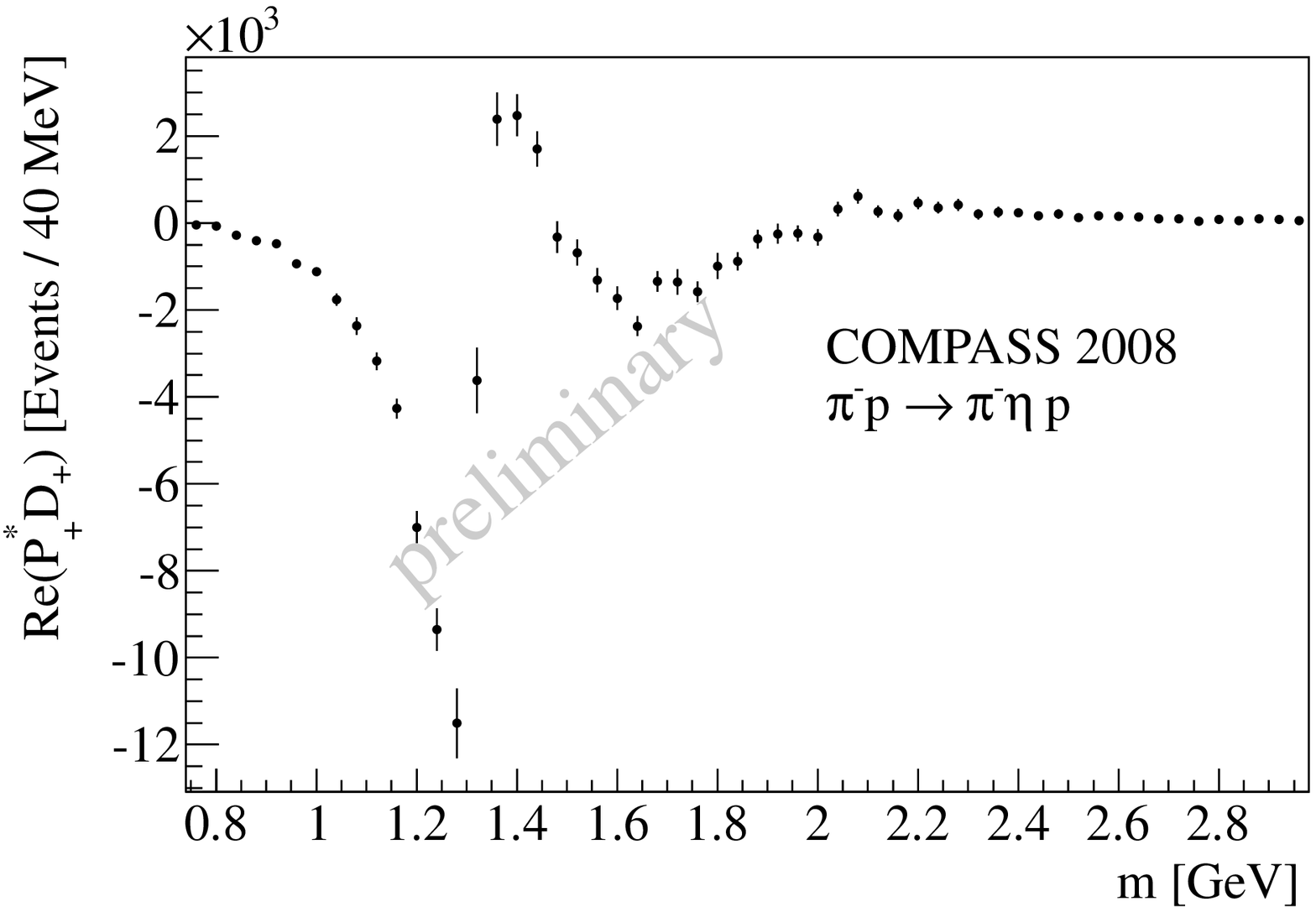}}
  \subfloat[$\textrm{Re}(P_{+}^*D_{++})$]{\includegraphics[width=.24\textwidth]{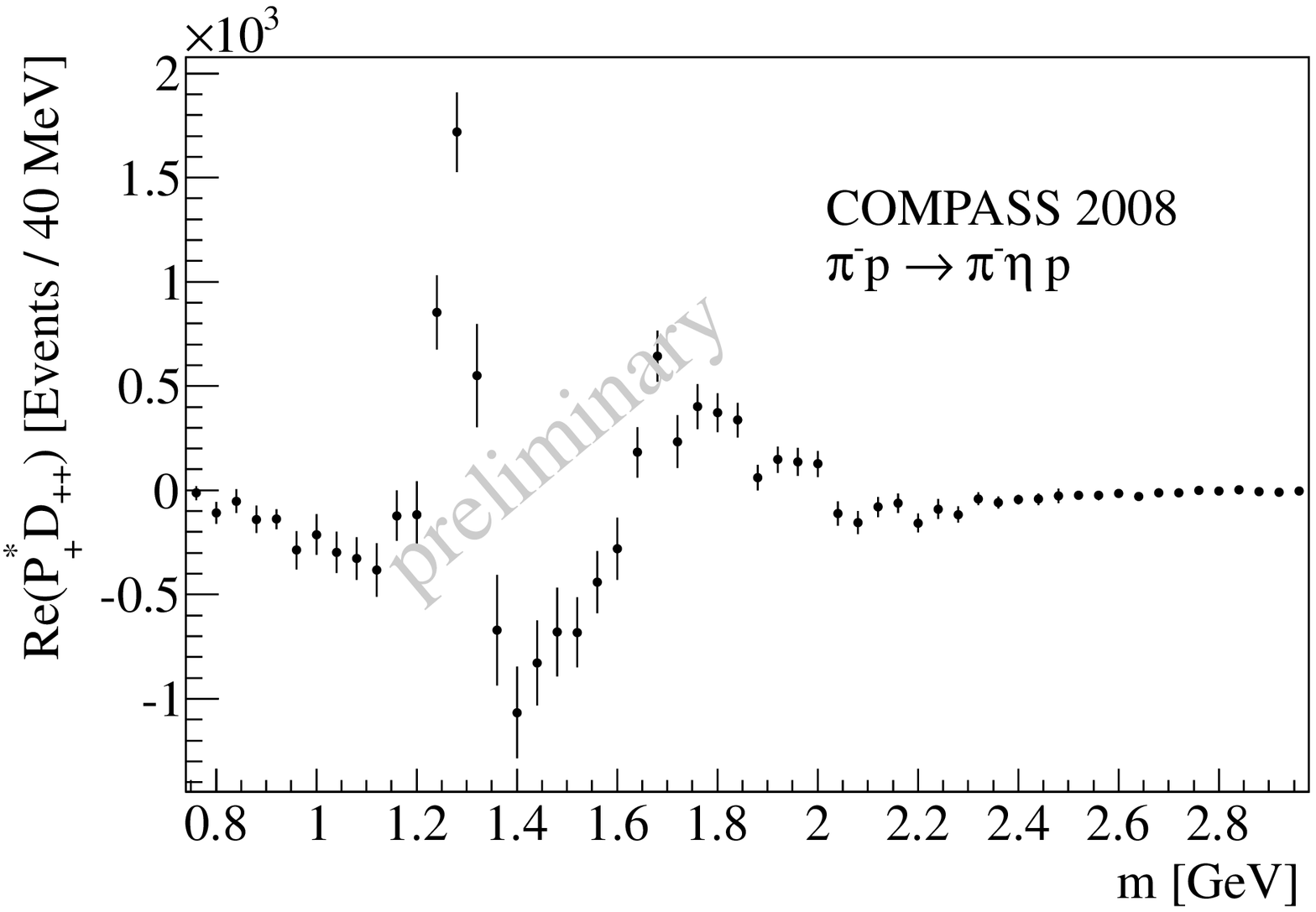}}
  \subfloat[$\textrm{Re}(P_{+}^*G_{+})$]{\includegraphics[width=.24\textwidth]{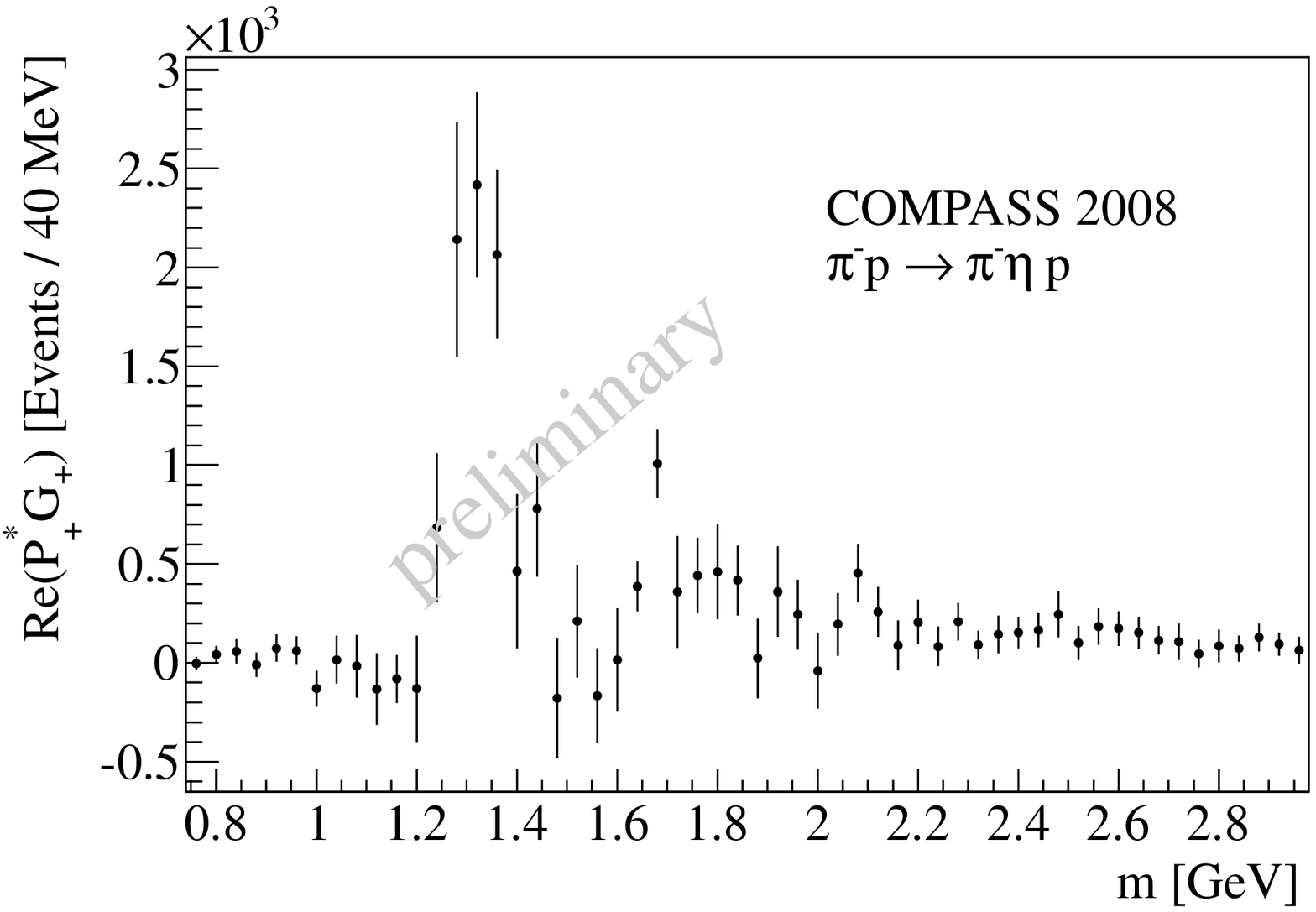}}\\

  \subfloat[$\textrm{Im}(P_{+}^*D_{+})$]{\includegraphics[width=.24\textwidth]{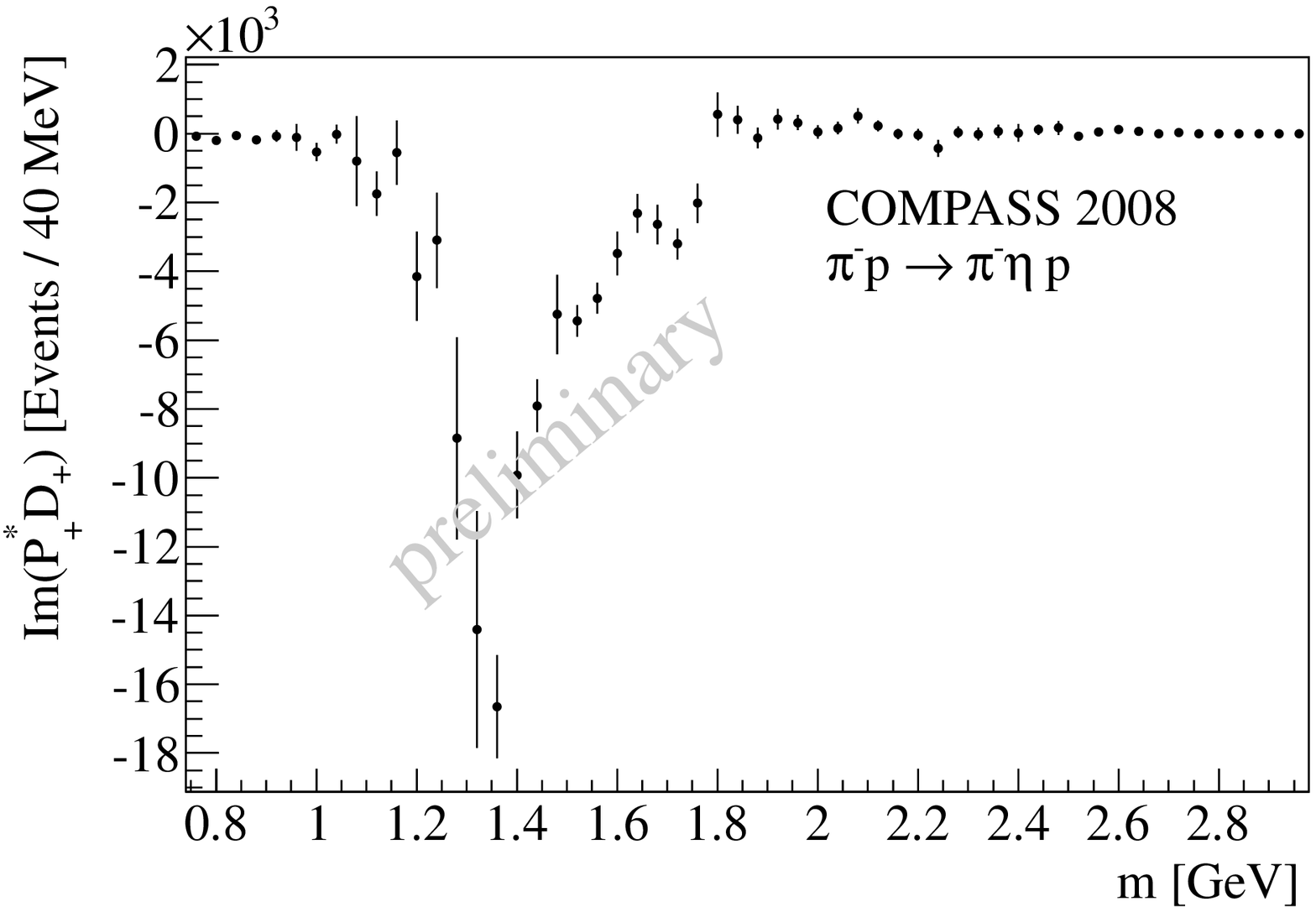}}
  \subfloat[$|D_+|^2$]{\includegraphics[width=.24\textwidth]{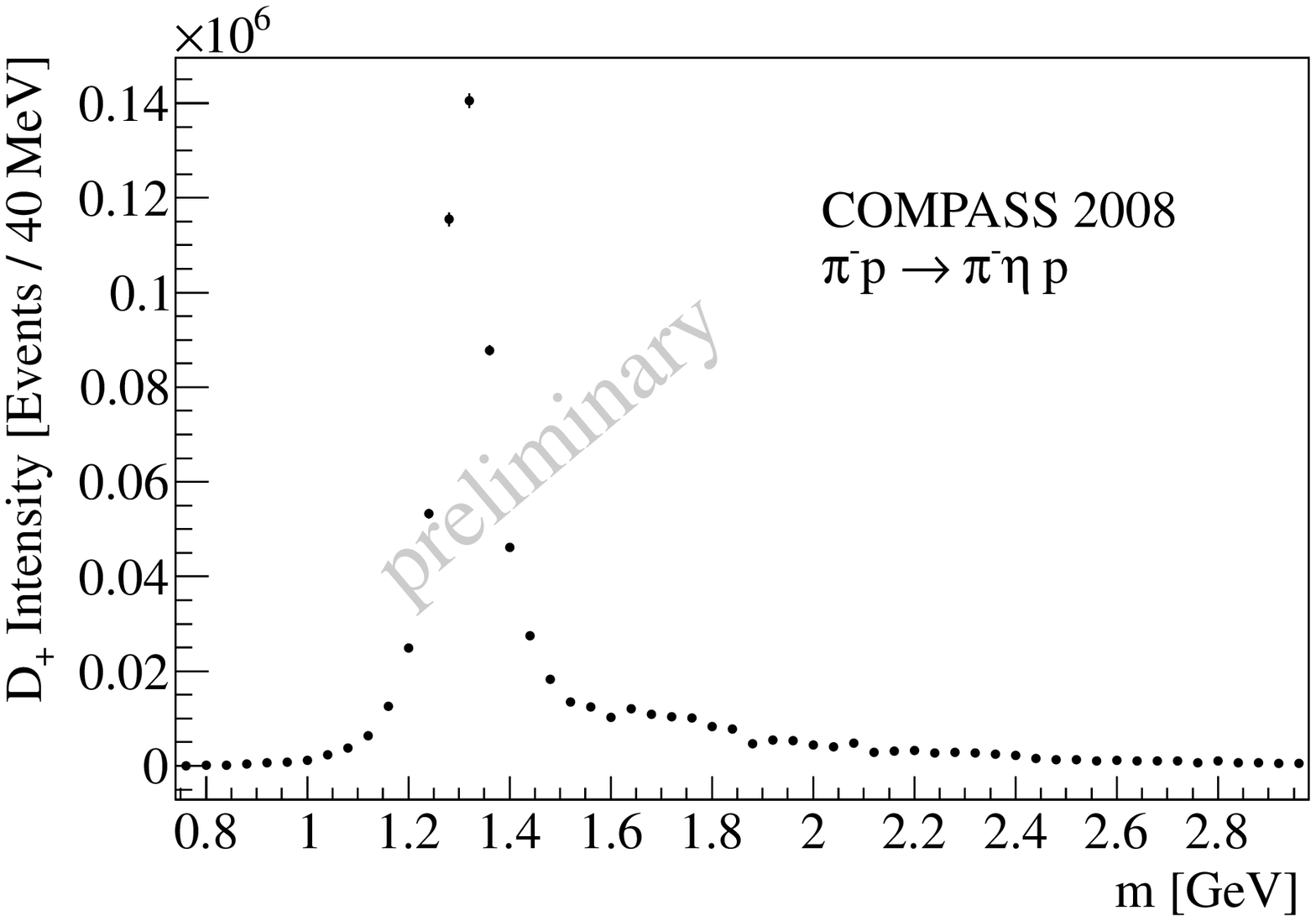}}
  \subfloat[$\textrm{Re}(D_{+}^*D_{++})$]{\includegraphics[width=.24\textwidth]{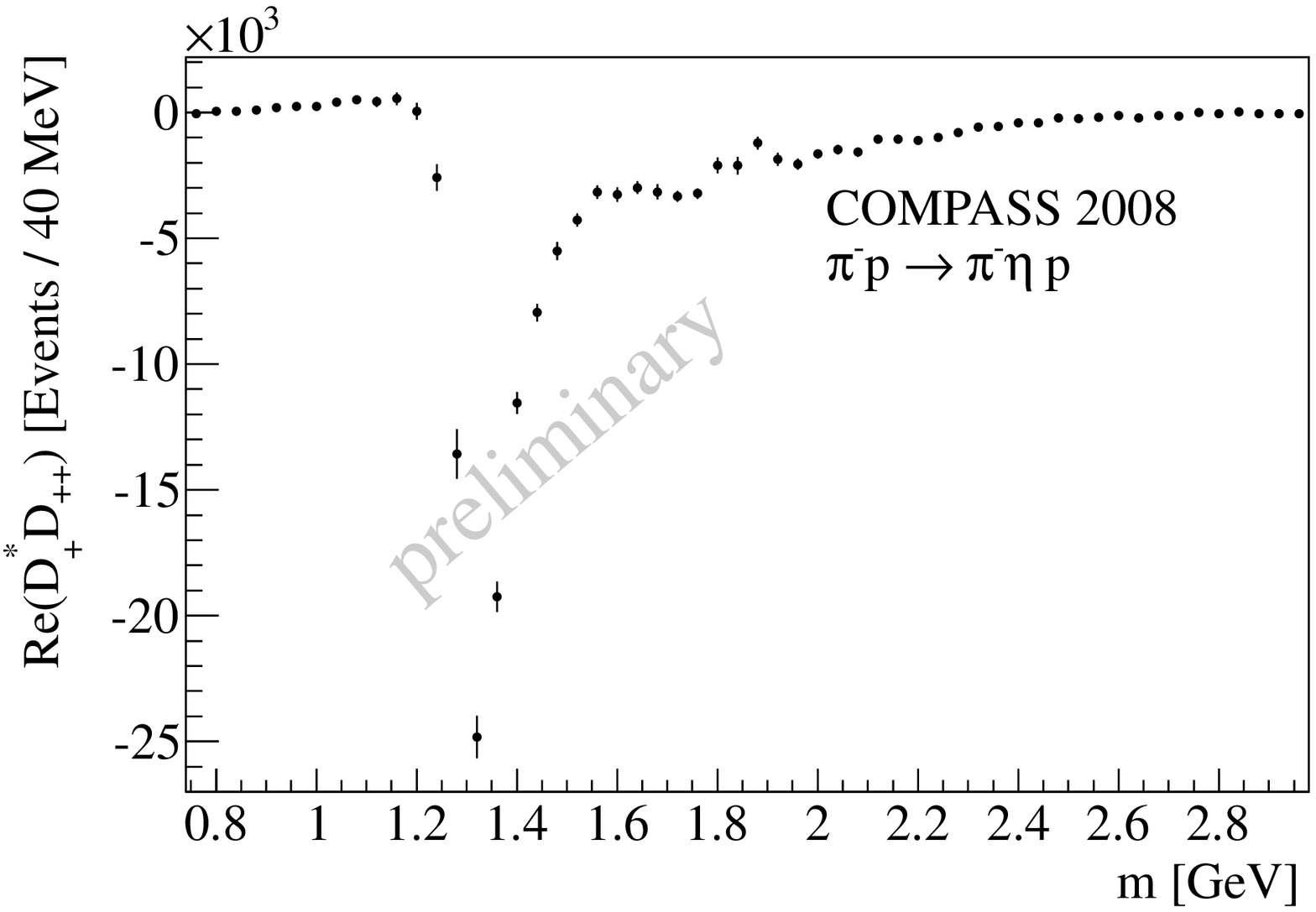}}
  \subfloat[$\textrm{Re}(D_{+}^*G_{+})$]{\includegraphics[width=.24\textwidth]{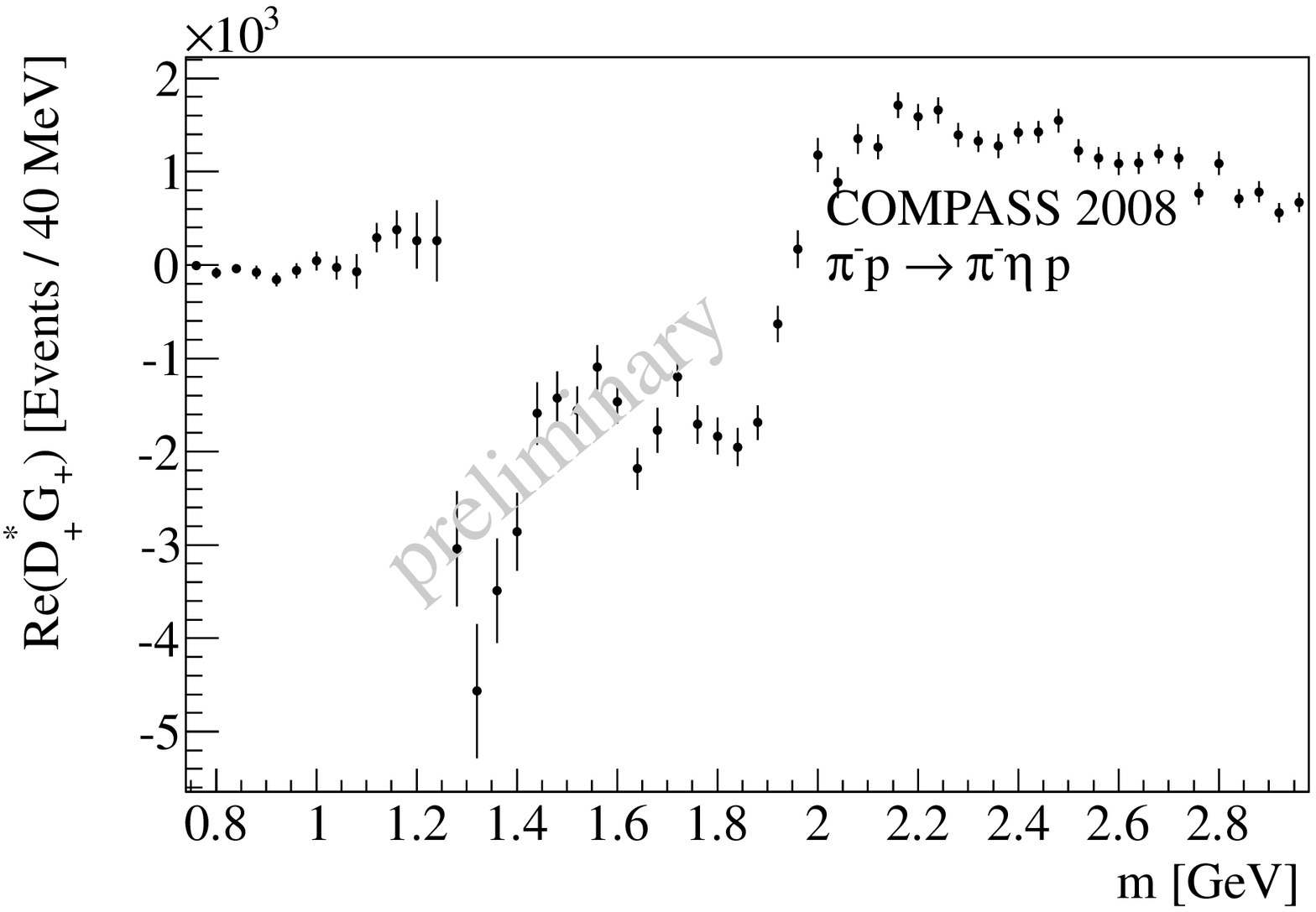}}\\

  \subfloat[$\textrm{Im}(P_{+}^*D_{++})$]{\includegraphics[width=.24\textwidth]{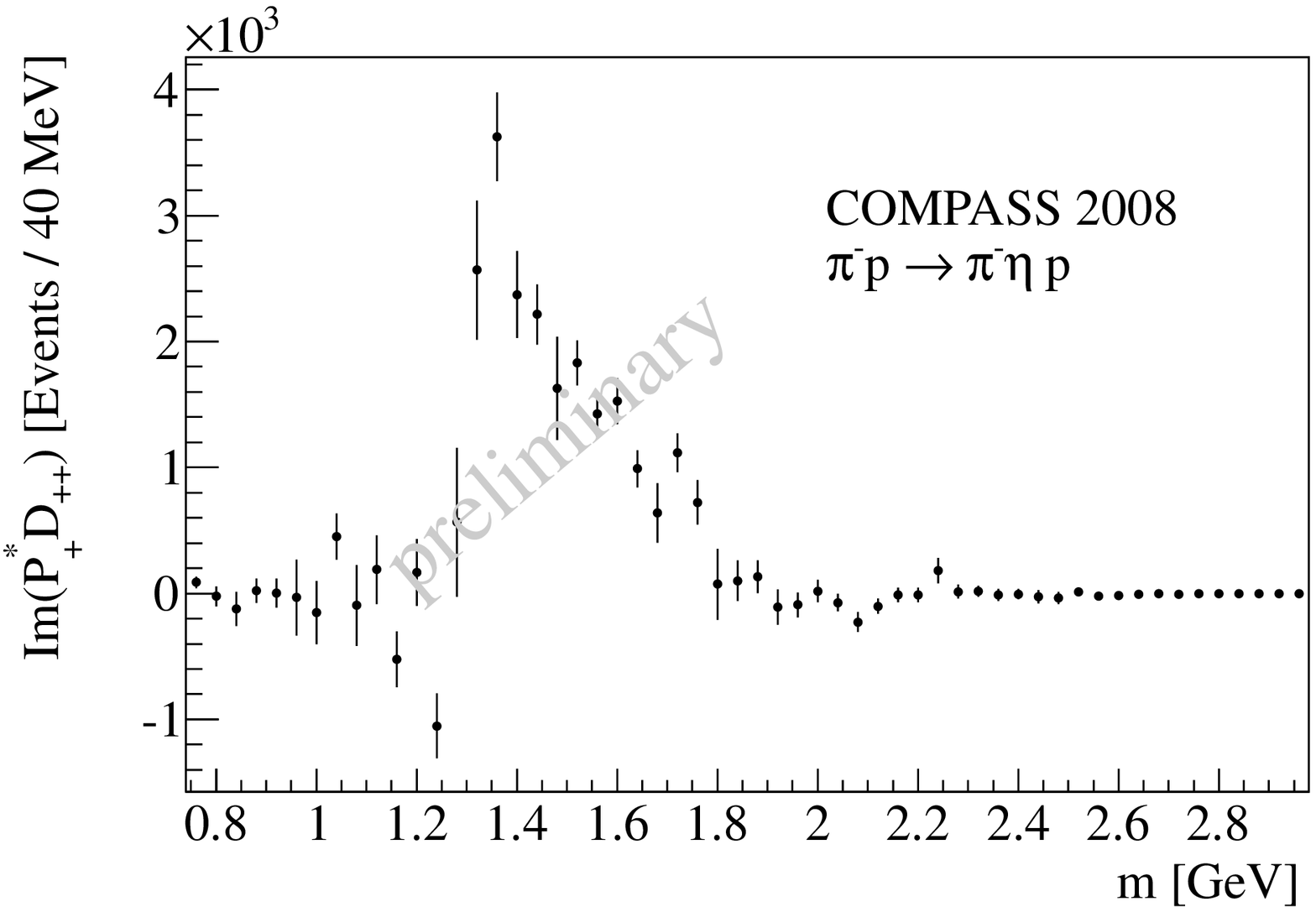}}
  \subfloat[$\textrm{Im}(D_{+}^*D_{++})$]{\includegraphics[width=.24\textwidth]{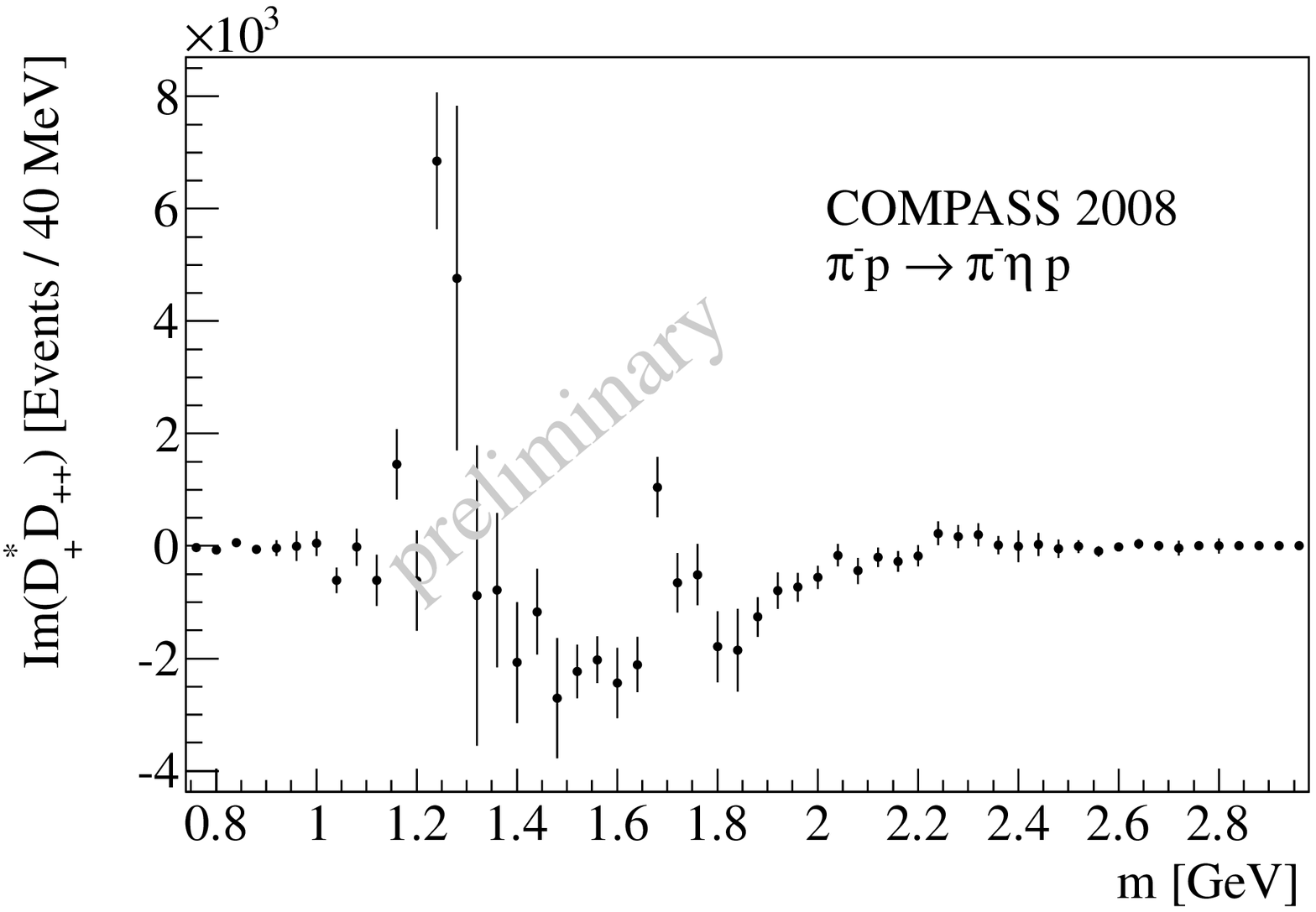}}
  \subfloat[$|D_{++}|^2$]{\includegraphics[width=.24\textwidth]{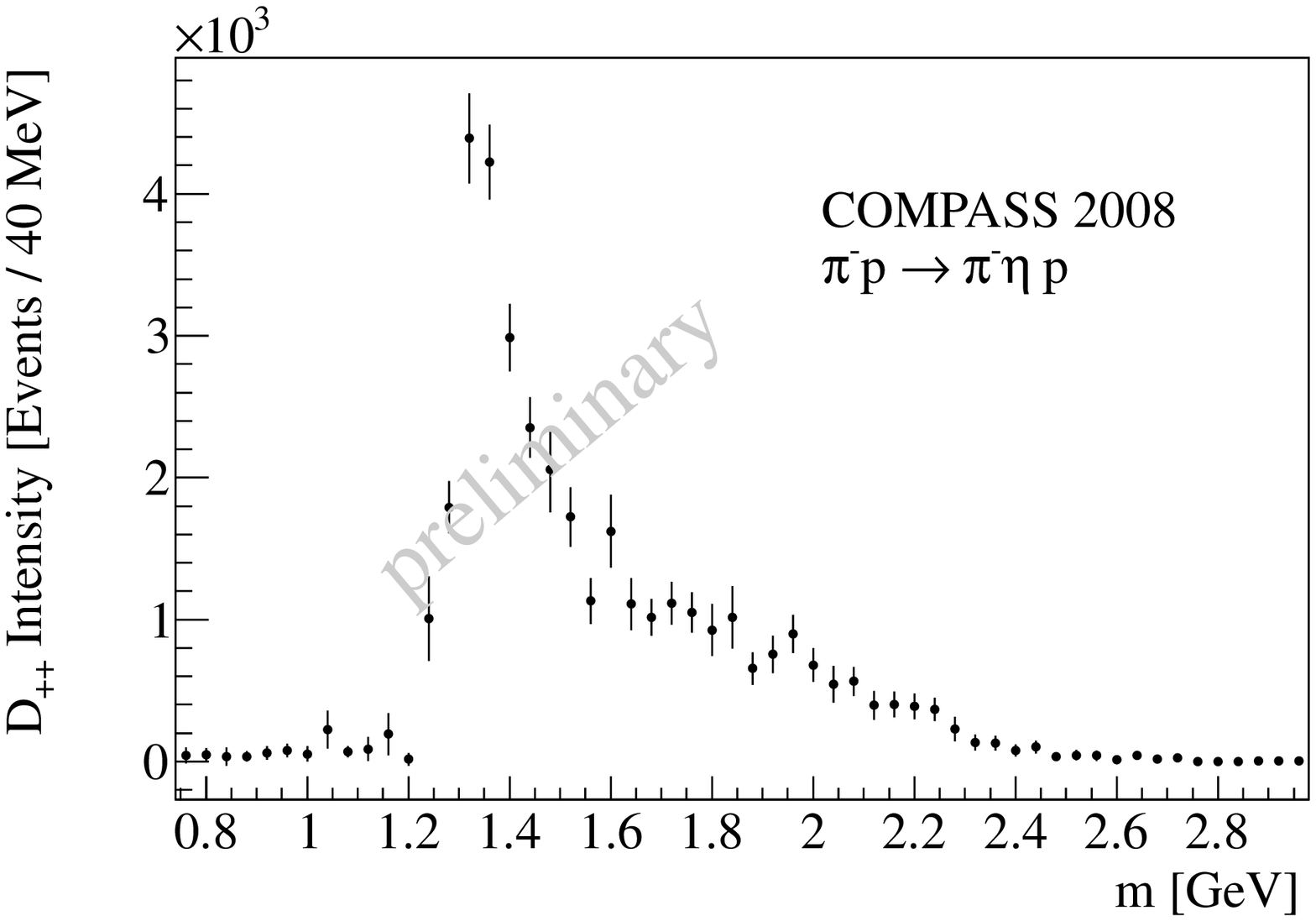}}
  \subfloat[$\textrm{Re}(G_{+}^*D_{++})$]{\includegraphics[width=.24\textwidth]{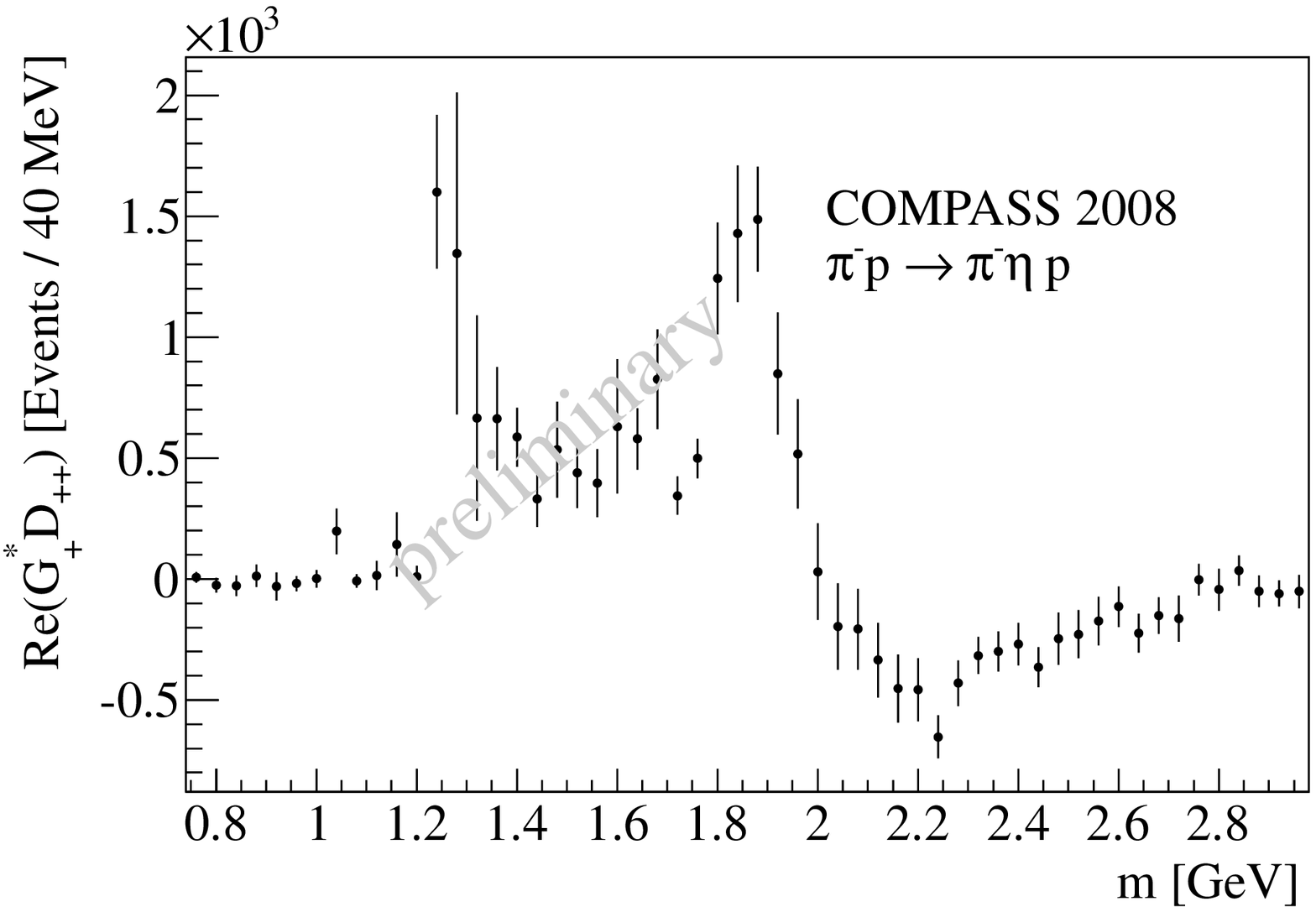}}\\

\subfloat[$\textrm{Im}(P_{+}^*G_{+})$]{\includegraphics[width=.24\textwidth]{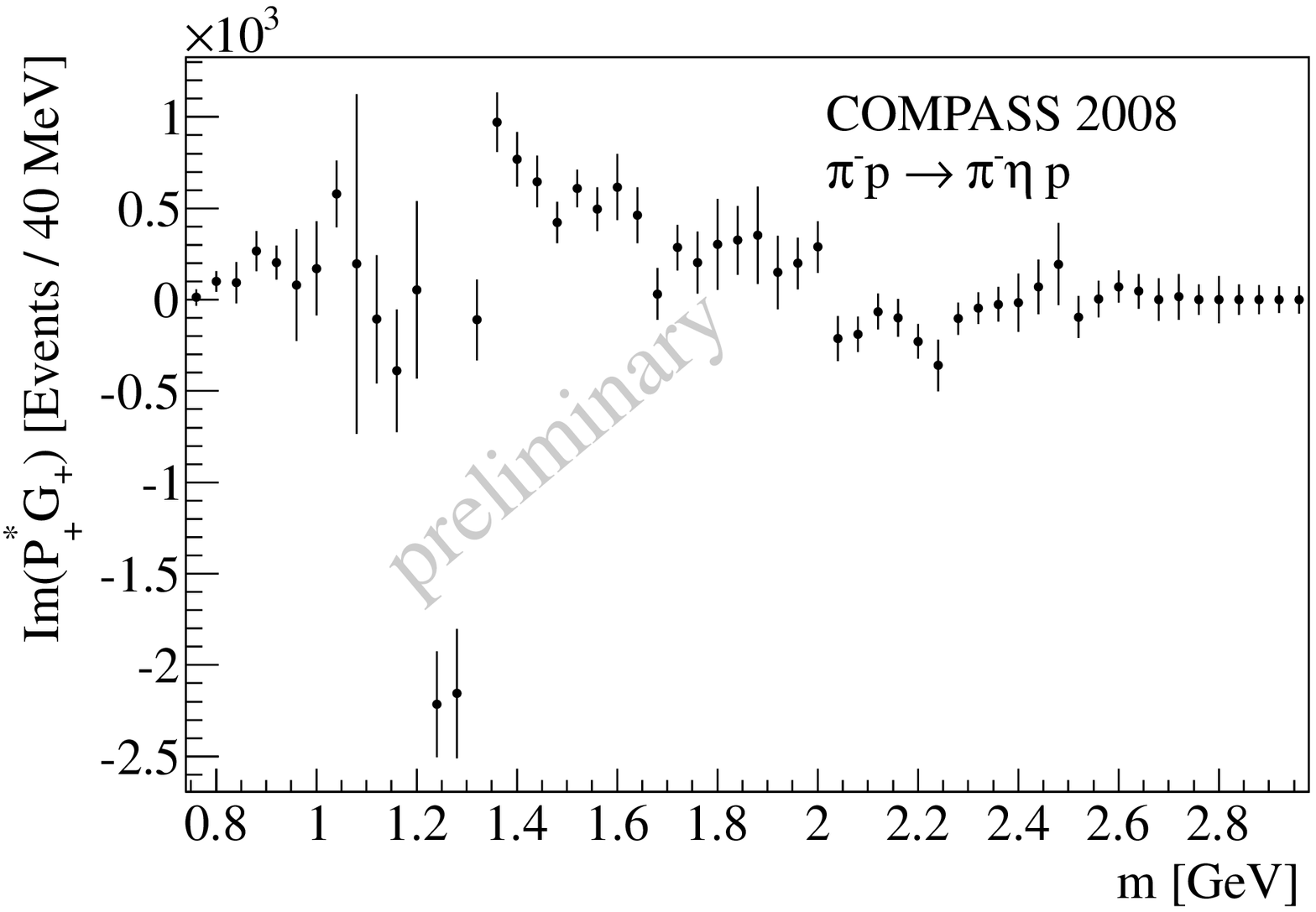}}
\subfloat[$\textrm{Im}(D_{+}^*G_{+})$]{\includegraphics[width=.24\textwidth]{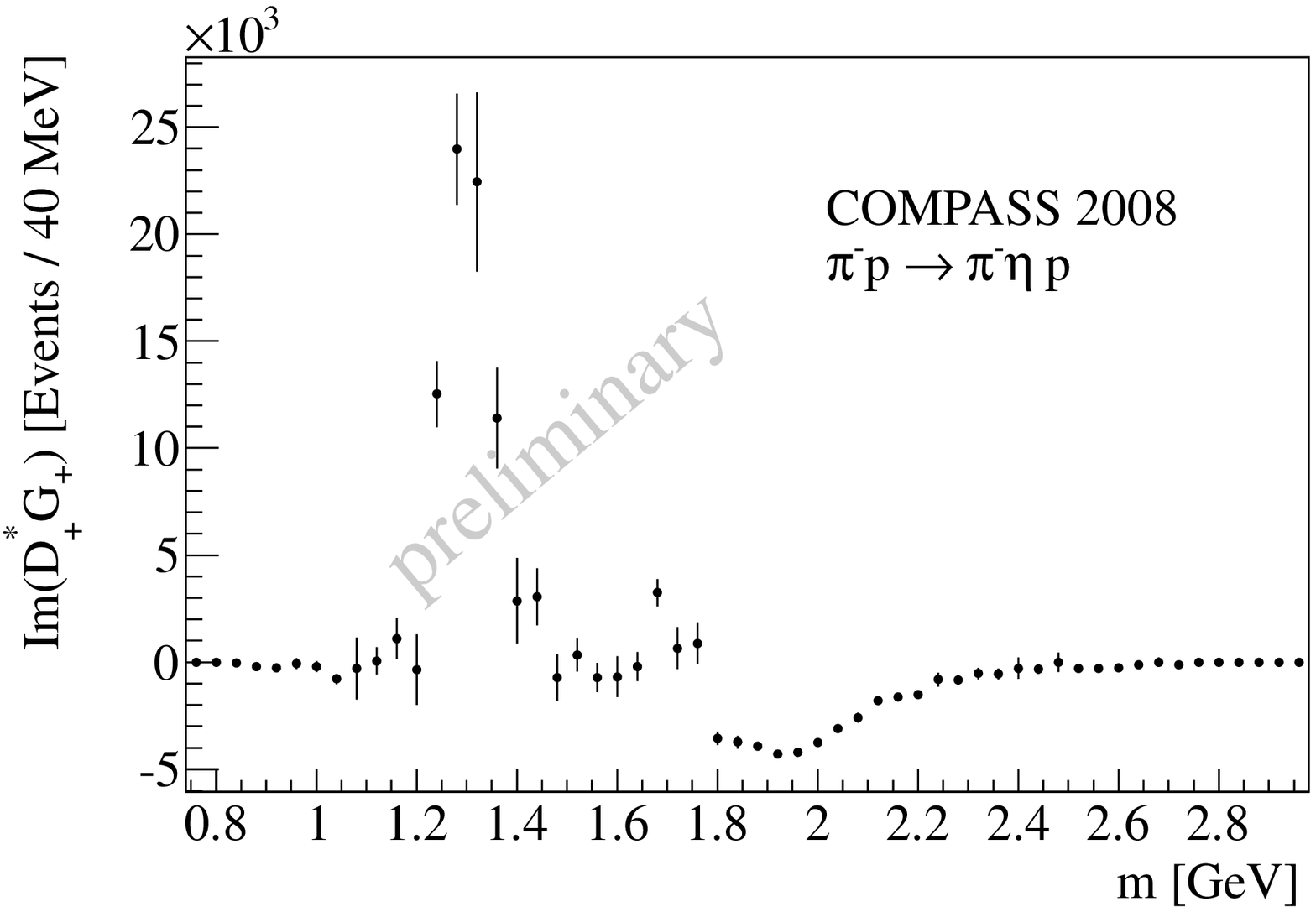}}
\subfloat[$\textrm{Im}(G_{+}^*D_{++})$]{\includegraphics[width=.24\textwidth]{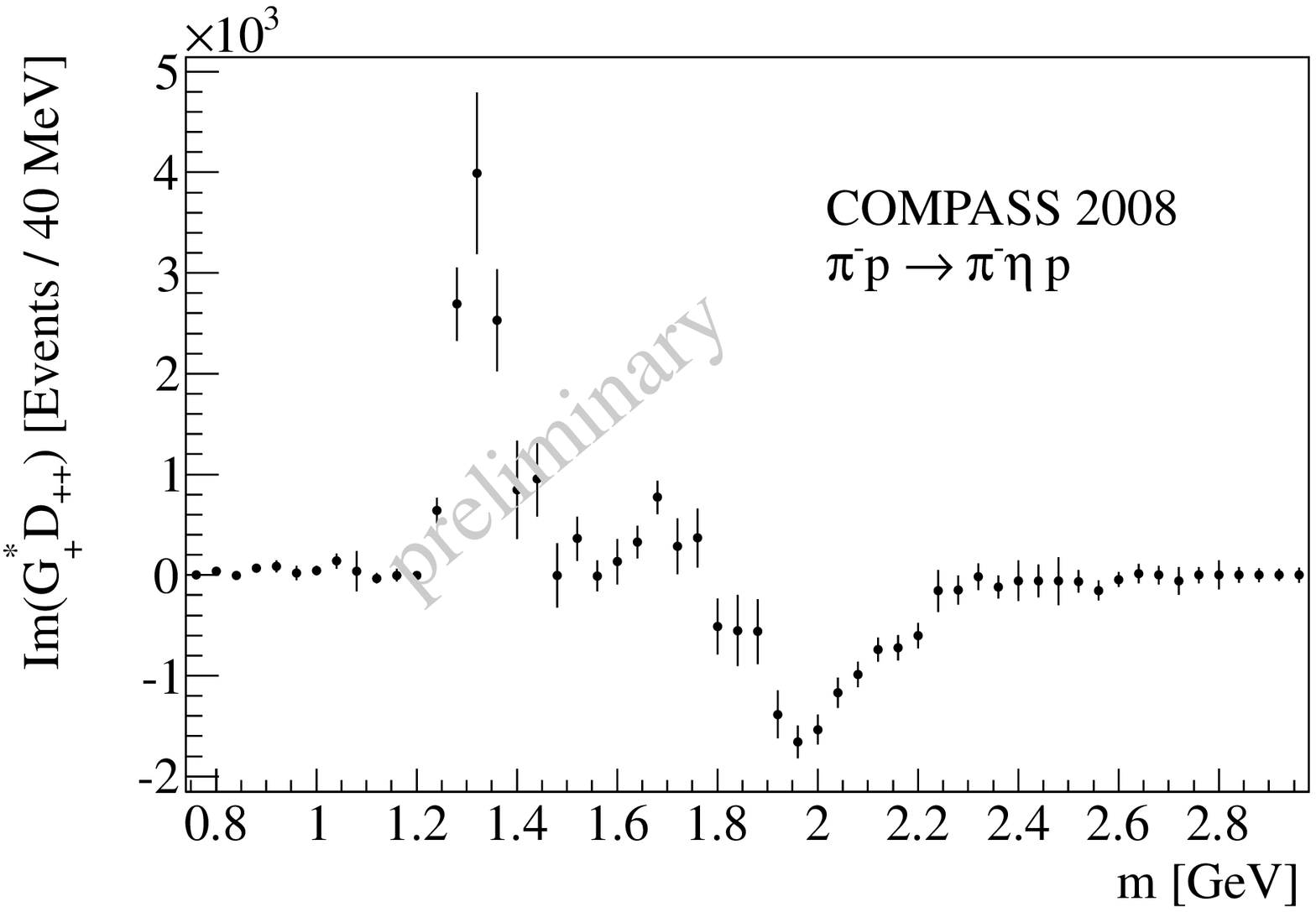}}
  \subfloat[$|G_+|^2$]{\includegraphics[width=.24\textwidth]{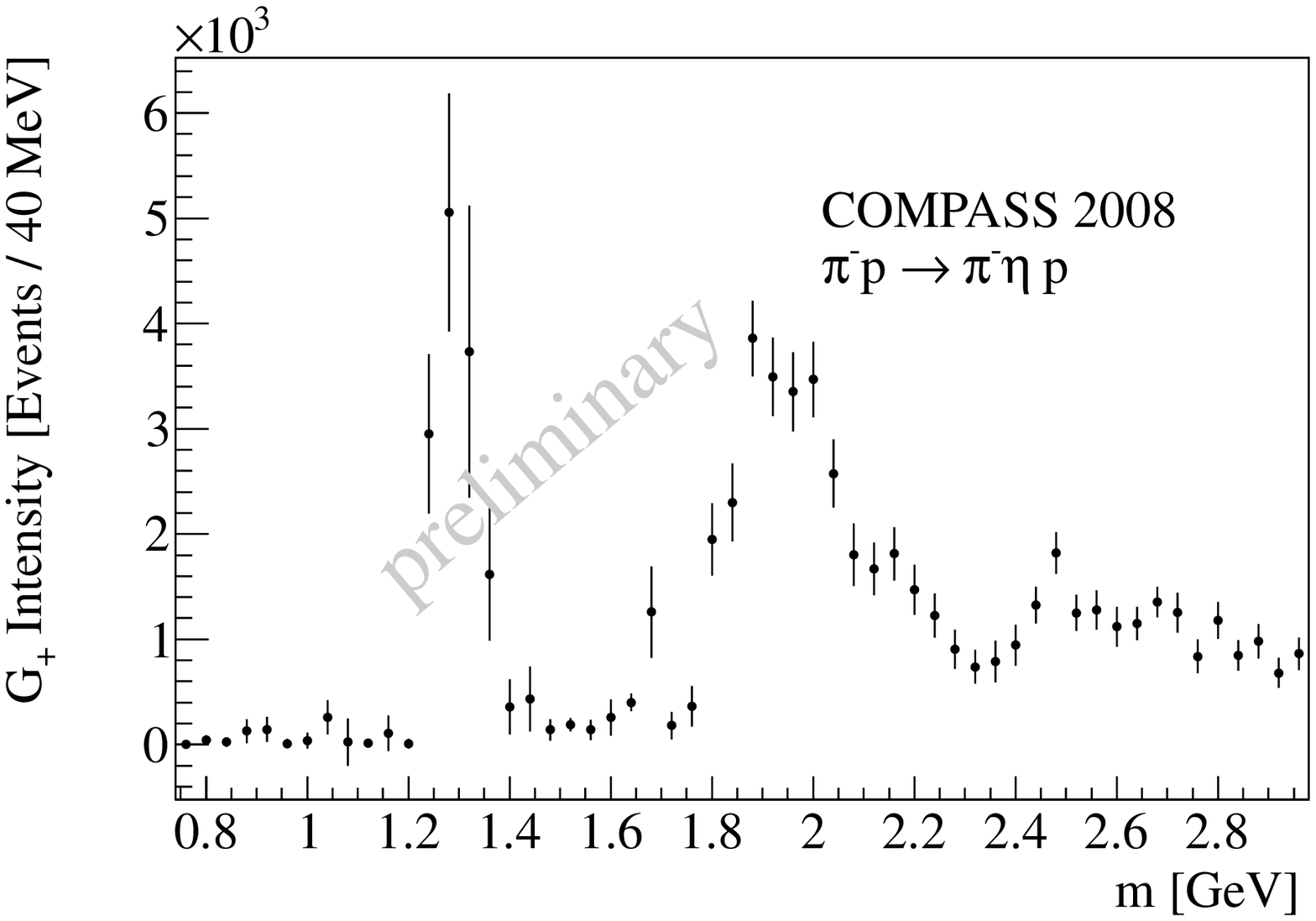}}

  \caption[Mass-independent partial-wave analysis of the $\pi^-\eta$
  system]{Mass-independent partial-wave analysis of the $\pi^-\eta$
    system.  The matrix shows on the diagonal the intensities of the
    natural-parity waves.  Above the diagonal are shown the respective
    relative real parts, below the respective relative imaginary
    parts.  The signs of the imaginary parts are not determined by the
    fit.  The dominating $D_+$ wave leaks into the $G_{++}$ wave in
    the mass range near $1.3\,\textrm{GeV}$.}
  \label{fig:massindep-etaPi}
\end{figure}

The analysis of the $\eta\pi^-$ data is performed in the same way as
was done for the $\eta'\pi^-$ data described in our previous report,
but due to the larger data set, we were able to add another wave,
namely the $m=2$ spin-2 $D_{++}$-wave.  This wave was previously
observed in interference terms extracted from the $\eta\pi^0$
system~\cite{Lednev:1997sa}.  We mention that unlike most previously
published analyses we also include the spin-4 $G_+$-wave.

Additional fits including natural-exchange spin-3, spin-5 and spin-6
waves were also performed, their presence being expected from a prior
analysis of the $K^-K^0_{{S}}$ system and double-Regge
phenomenology~\cite{Martin:1978jn,Shimada:1978sx}.  With these waves
included, the data can be described without recourse to
unnatural-exchange waves all the way up to $3\,\textrm{GeV}$, in
accordance with the expected dominance of the spin-parity natural
Pomeron exchange.  Since the inclusion of these waves leads to
mathematical ambiguities~\cite{Chung:1997qd}, and since the data in
the resonance-dominated range up to approx. $2\,\textrm{GeV}$ is
well-described with the smaller set, we have omitted them in the
depicted fits.

The fit results for the $\eta\pi^-$ data are shown in
Figs.~\ref{fig:massindep-etaPi} and the relative phases in red in
Fig.~\ref{fig:comparison-phases}.  Only the intensities and relative
real parts can be extracted by the fit, this leaves an ambiguity in
the sign of the imaginary part, which can in turn lead to
discontinuities and jumps in the calculated phases.  Additionally,
interpretation of these fits comes with the caveat that a continuous
ambiguity prevents the fit from accounting for incoherent
contributions, the phases therefore cannot be interpreted without
care~\cite{Martin:1978jn}.  Our data show a significant $P_+$ wave
which interferes with the dominant $D_+$ wave.  The size of the
$D_{++}$ wave relative to the $D_+$ wave is consistent with other
COMPASS analyses~\cite{Nerling:2011xs}.  Phase-motion due to the
$a_2(1320)$ and $a_4(2040)$ resonances can be clearly seen.  The
relative phase motion of the $D_+$ and $P_+$ waves is consistent with
previous analyses.

\section{Comparison of the Systems $\eta\pi^-$ and $\eta'\pi^-$}

\begin{figure}[htbp]
  \centering
  
\subfloat[$|P_+|^2$]{\includegraphics[width=.32\textwidth]{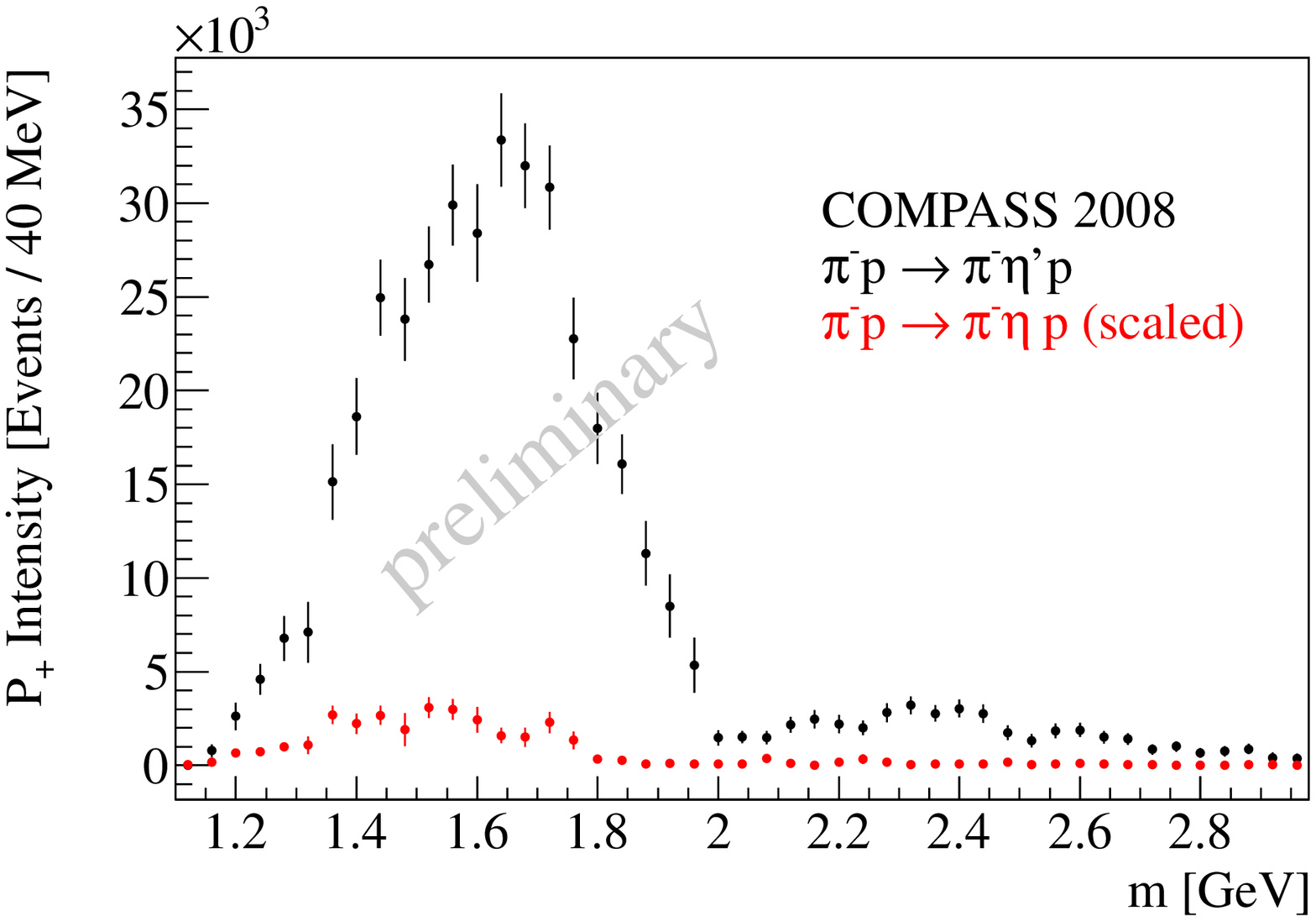}}
\subfloat[$\textrm{Re}(P_+^*D_+)$]{\includegraphics[width=.32\textwidth]{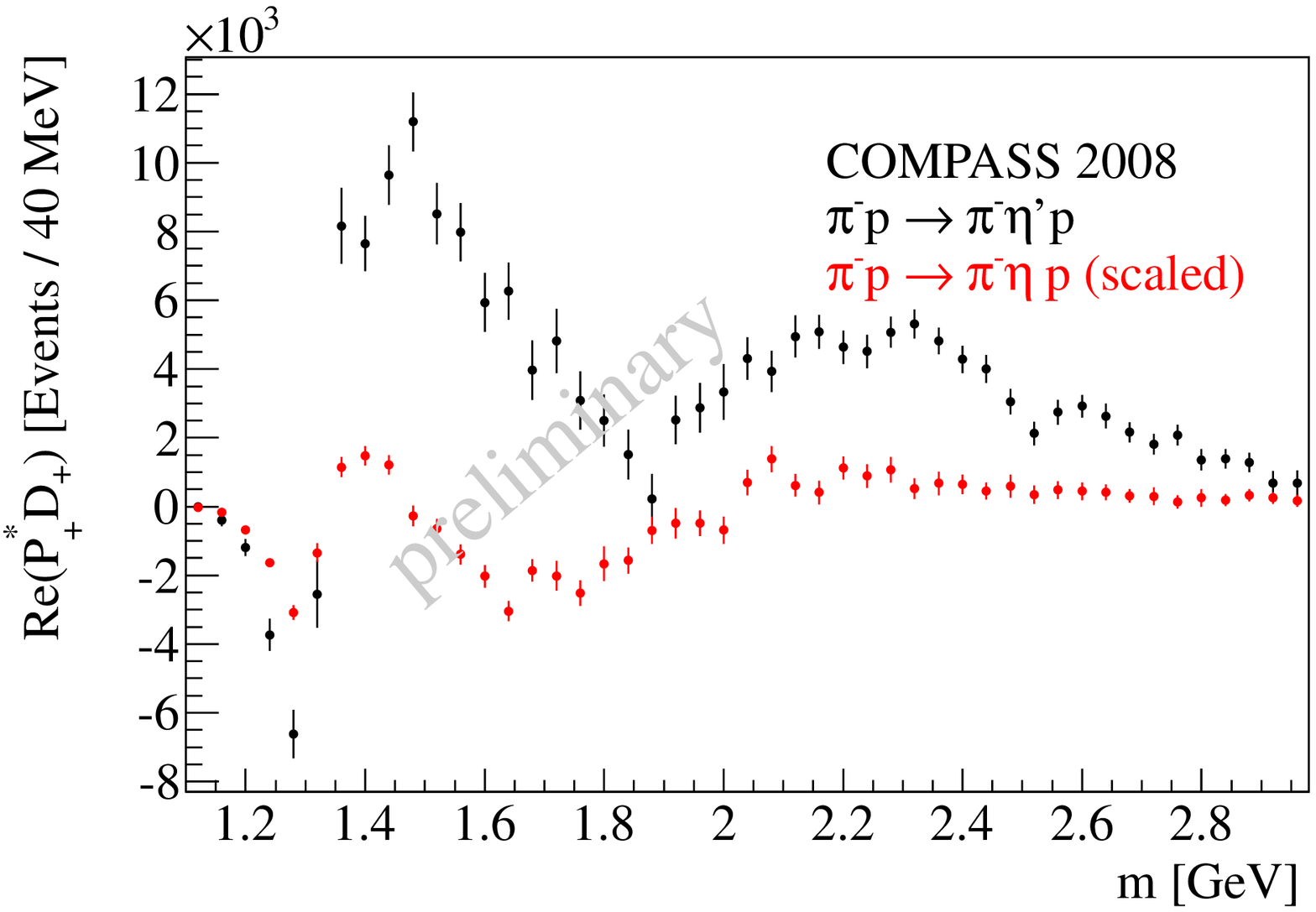}}
\subfloat[$\textrm{Re}(P_+^*G_+)$]{\includegraphics[width=.32\textwidth]{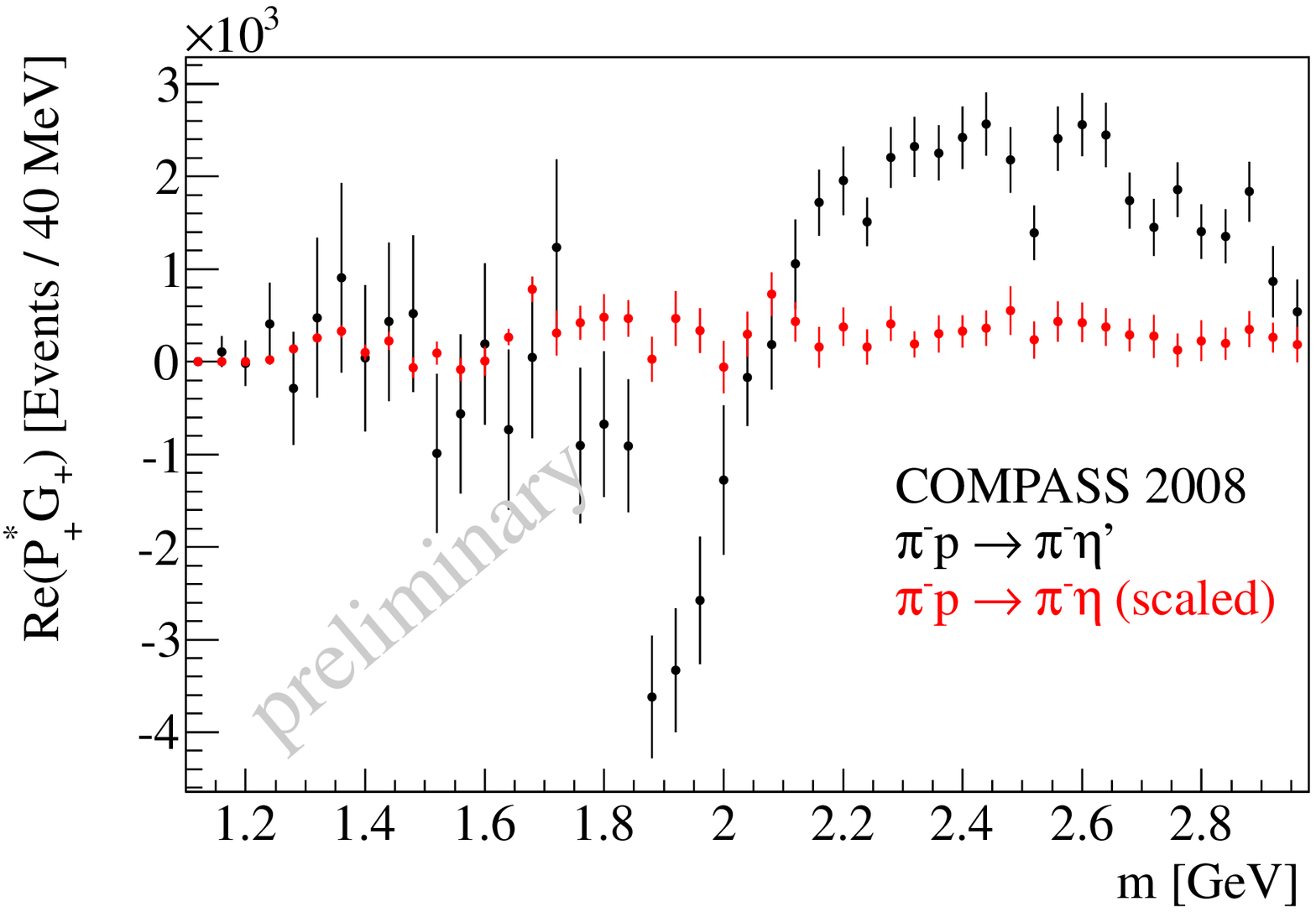}}\\
\subfloat[$\textrm{Im}(P_+^*D_+)$]{\includegraphics[width=.32\textwidth]{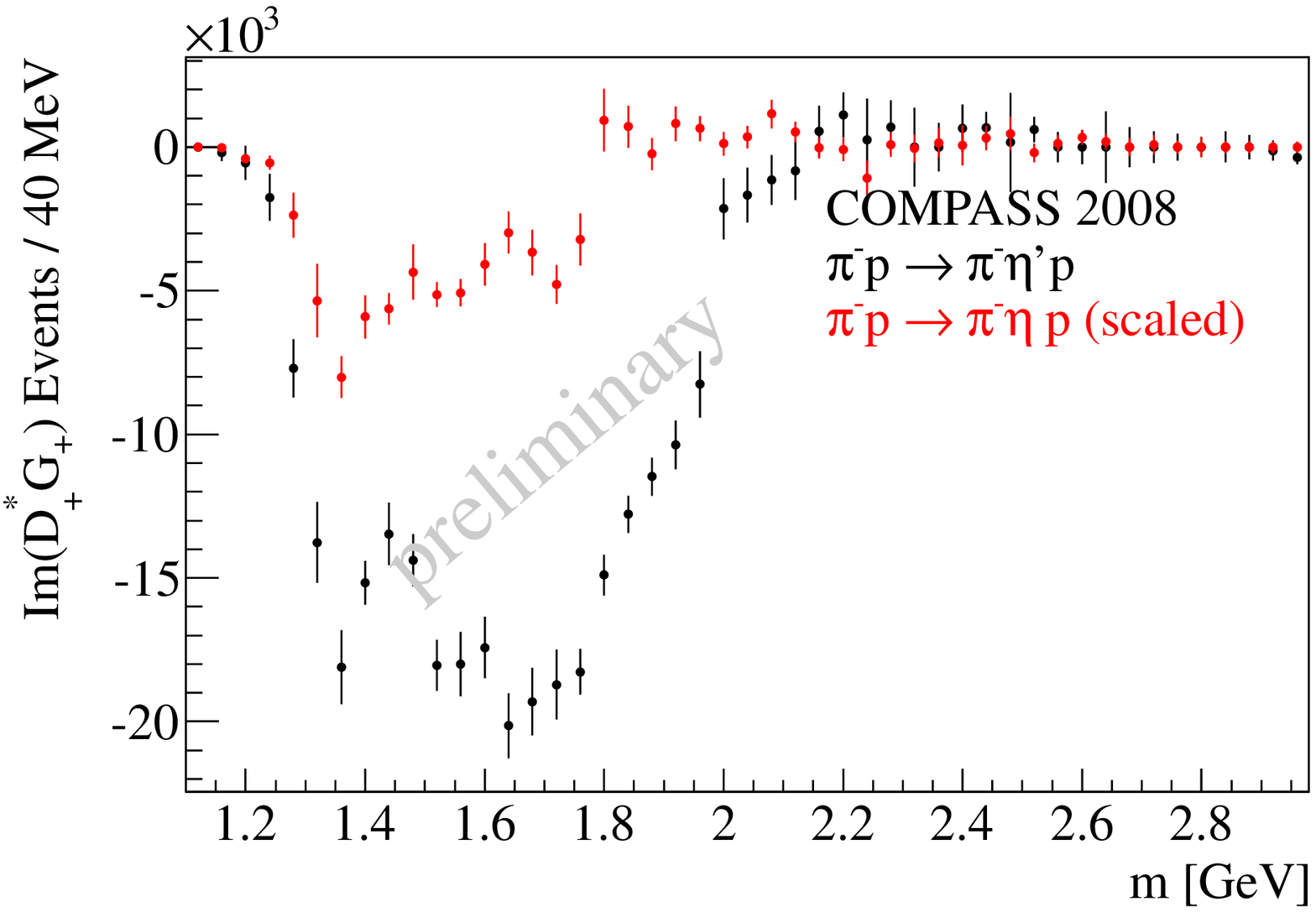}}
\subfloat[$|D_+|^2$]{\includegraphics[width=.32\textwidth]{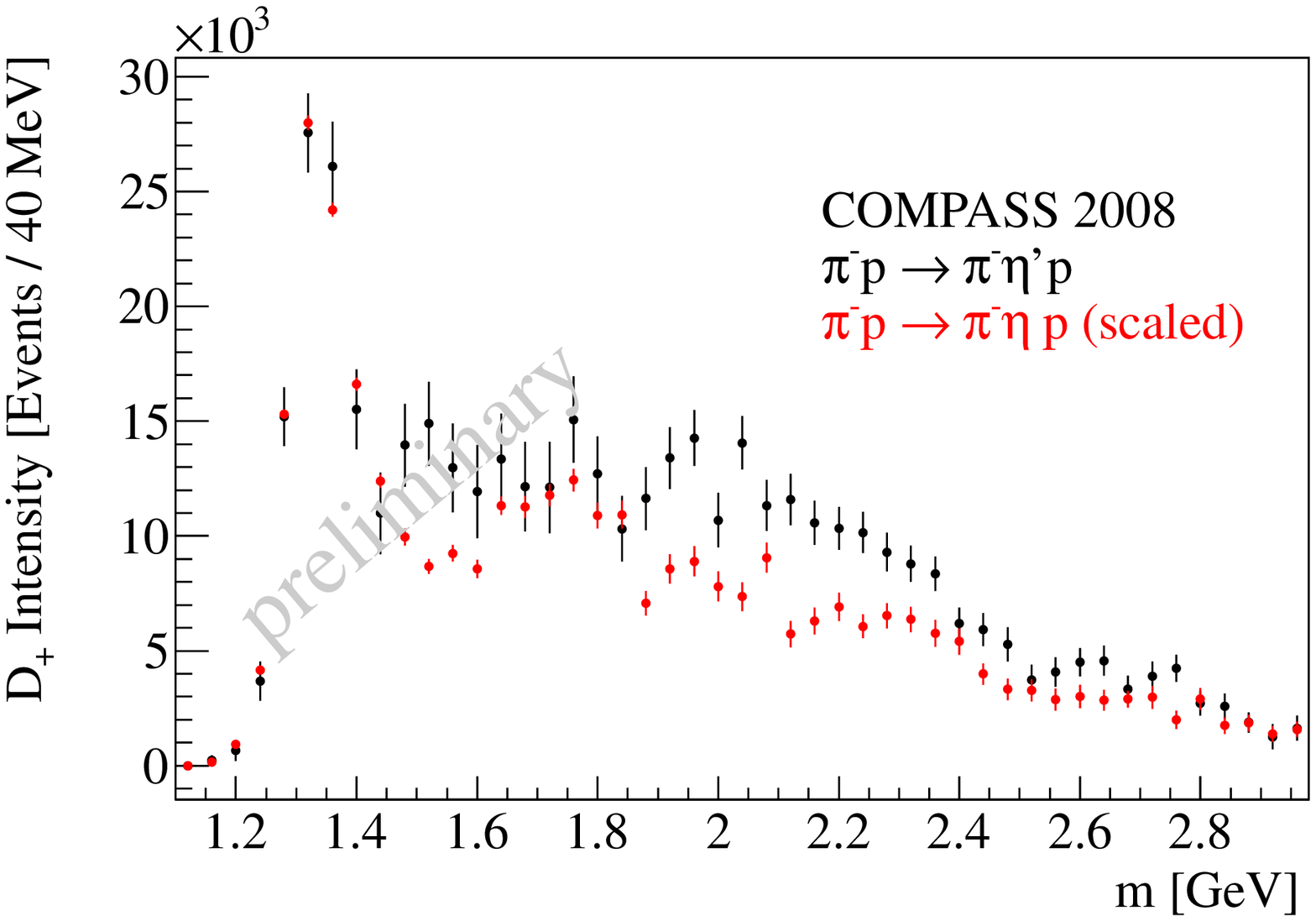}}
\subfloat[$\textrm{Re}(D_+^*G_+)$]{\includegraphics[width=.32\textwidth]{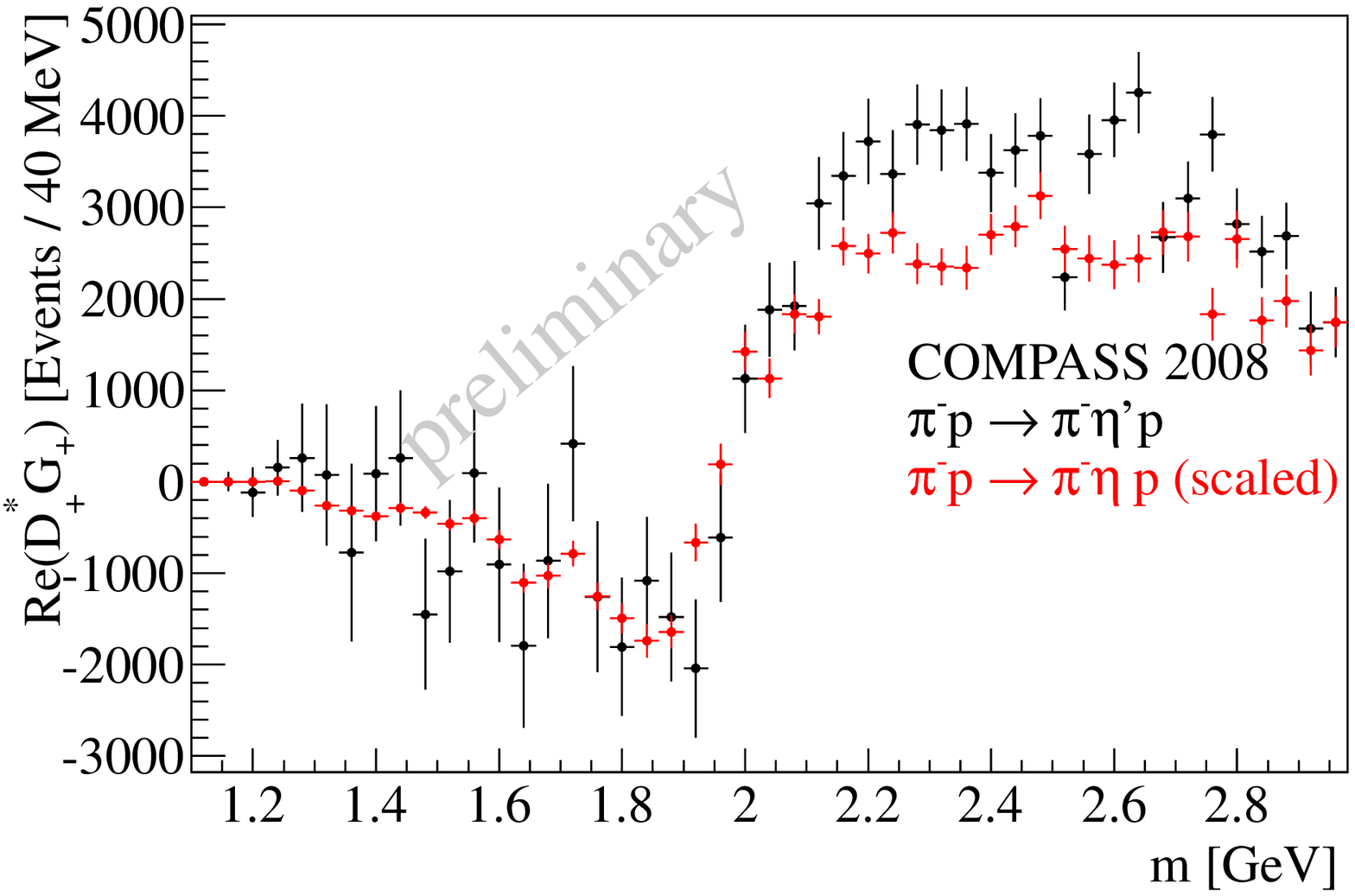}}\\
\subfloat[$\textrm{Im}(P_+^*G_+)$]{\includegraphics[width=.32\textwidth]{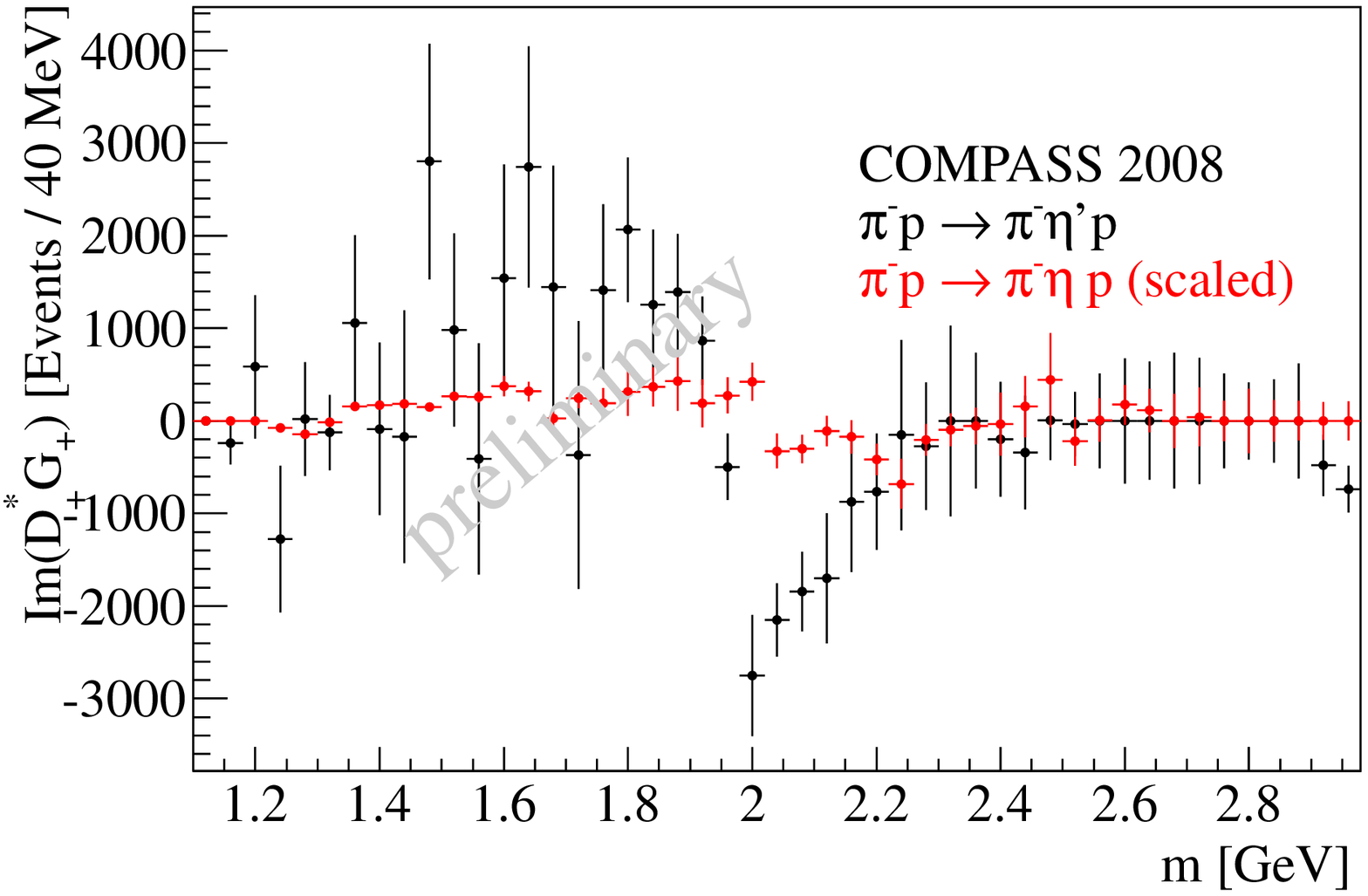}}
\subfloat[$\textrm{Im}(D_+^*G_+)$]{\includegraphics[width=.32\textwidth]{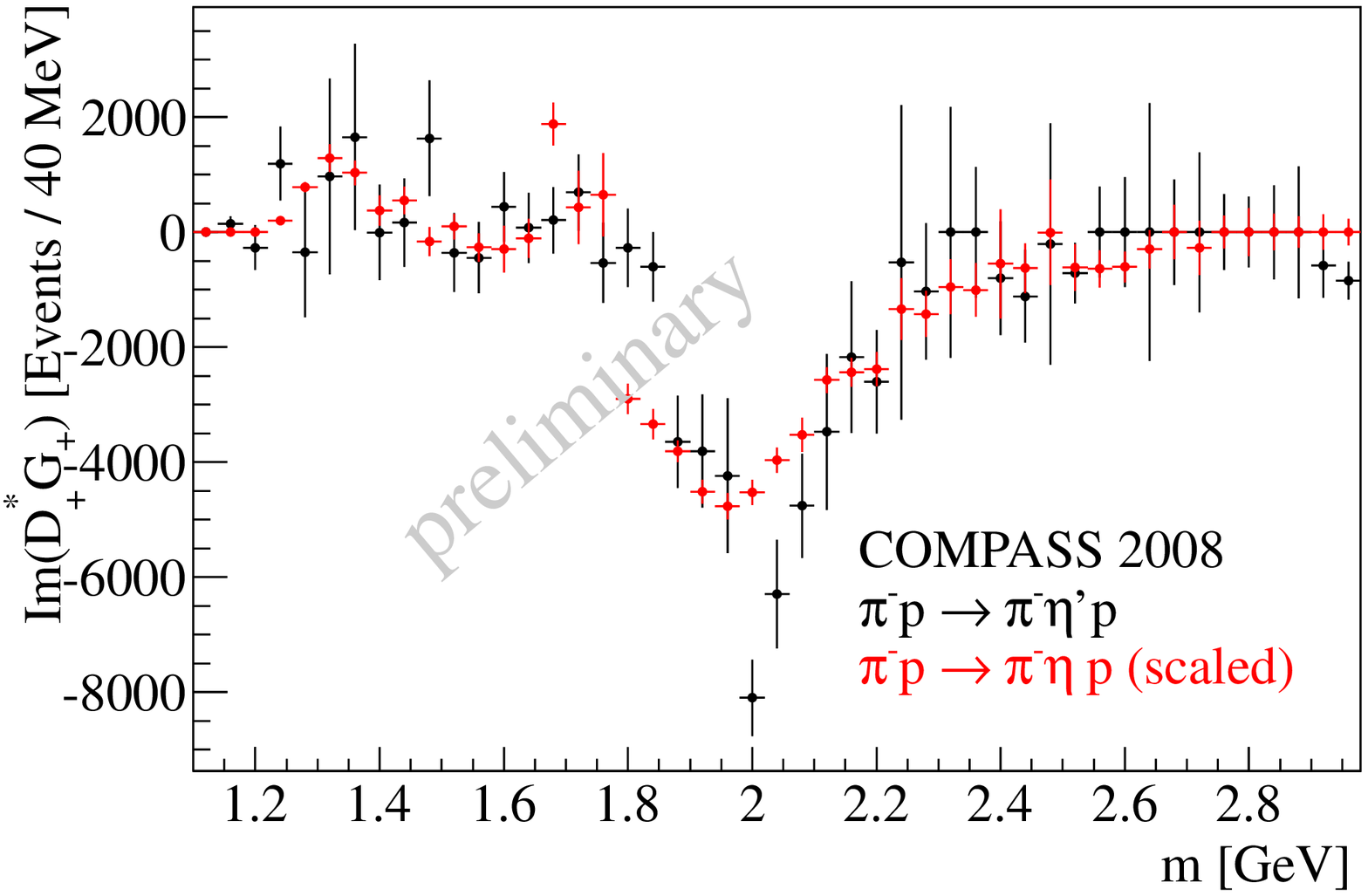}}
\subfloat[$|G_+|^2$]{\includegraphics[width=.32\textwidth]{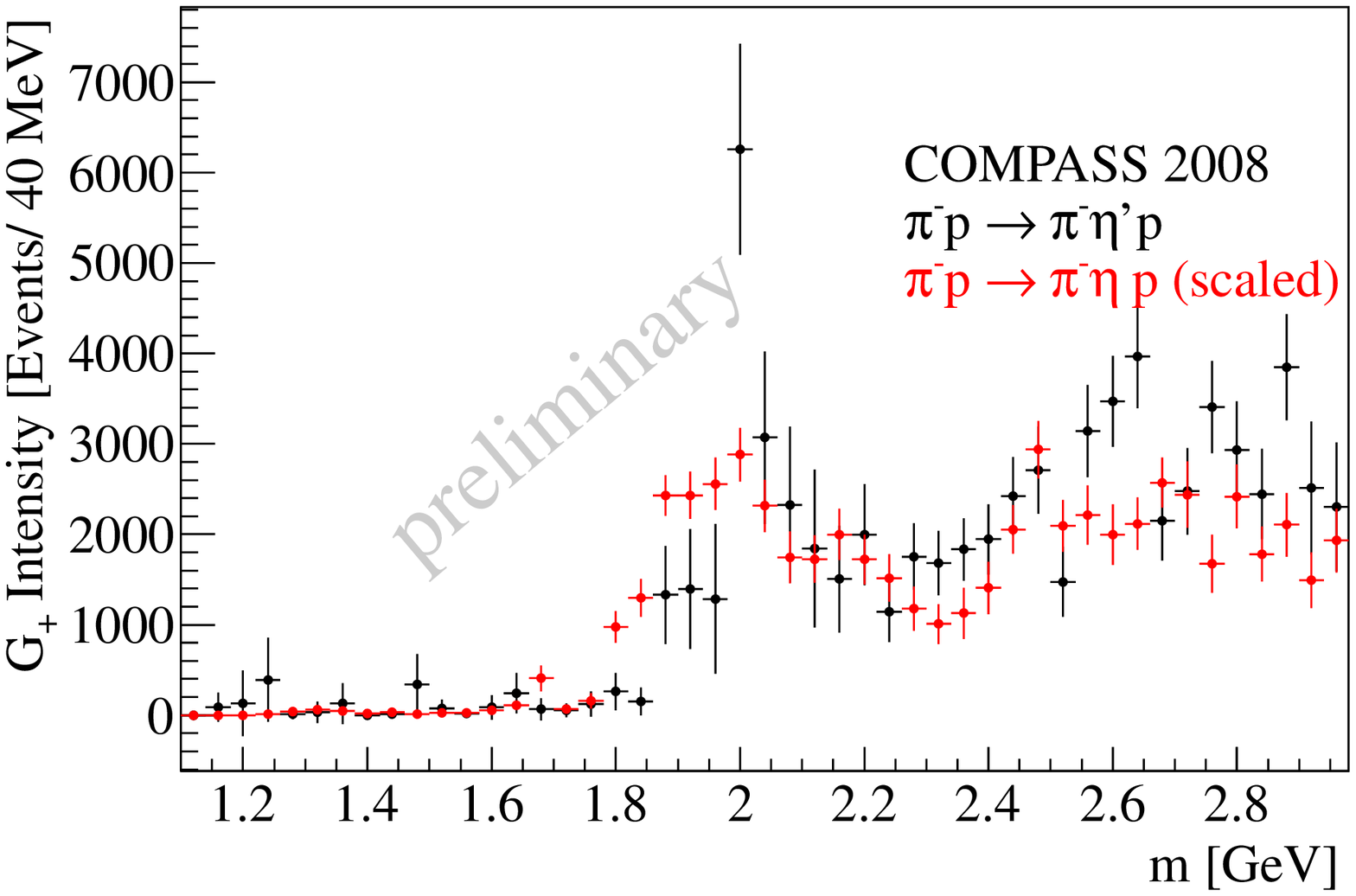}}
\caption[Comparison of $\eta'\pi$ and $\eta\pi$]{Comparison of the
  partial-wave amplitudes obtained in the $\pi\eta'$ (black) and
  $\pi\eta$ systems (red) after re-scaling with the phase-space
  factors.}
  \label{fig:comparison}
\end{figure}

The physical $\eta$ and $\eta'$ mesons are not independent objects but
mixtures of the $SU(3)$ flavor basis states $\eta_s=s\overline s$ and
$\eta_n=u\overline u + d\overline d$.  As such, the relative strength
of their production can be expressed in terms of the mixing angle
$\phi$ and phase-space and dynamical (barrier)
factors~\cite{Okubo:1976gi}.  Taking the simplest form for the
dynamical factor that yields the correct asymptotic behavior near
threshold, $F_J(q)=q^J$ ($q^{(\prime)}$ the breakup momentum into
$\eta^{(\prime)}\pi$ at the given invariant mass), and taking into
account phase-space, we rescale the $\eta\pi^-$ amplitudes with the
factor $(q'/q)^{J+1/2}$ and overlay them on the $\eta'\pi^-$
amplitudes.  The resultant matrix of overlaid fit results (omitting
the $D_{++}$ not included in $\eta'\pi^-$) is shown in
Fig.~\ref{fig:comparison}.

\begin{figure}[tbp]
  \centering
  \subfloat[$\arg(D_+/P_+)$]%
  {\includegraphics[width=.4\textwidth]{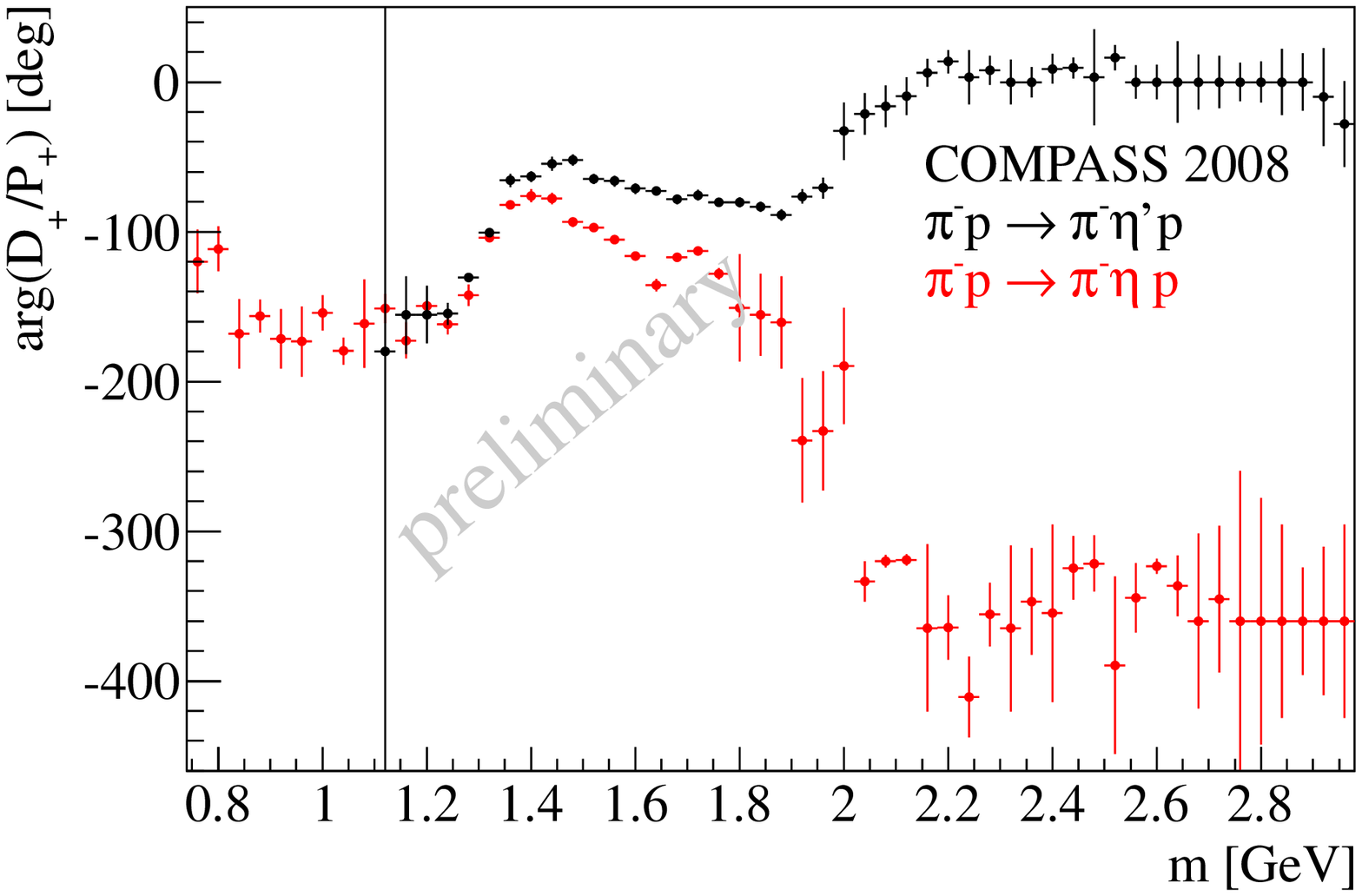}}
  \subfloat[$\arg(G_+/D_+)$]%
  {\includegraphics[width=.4\textwidth]{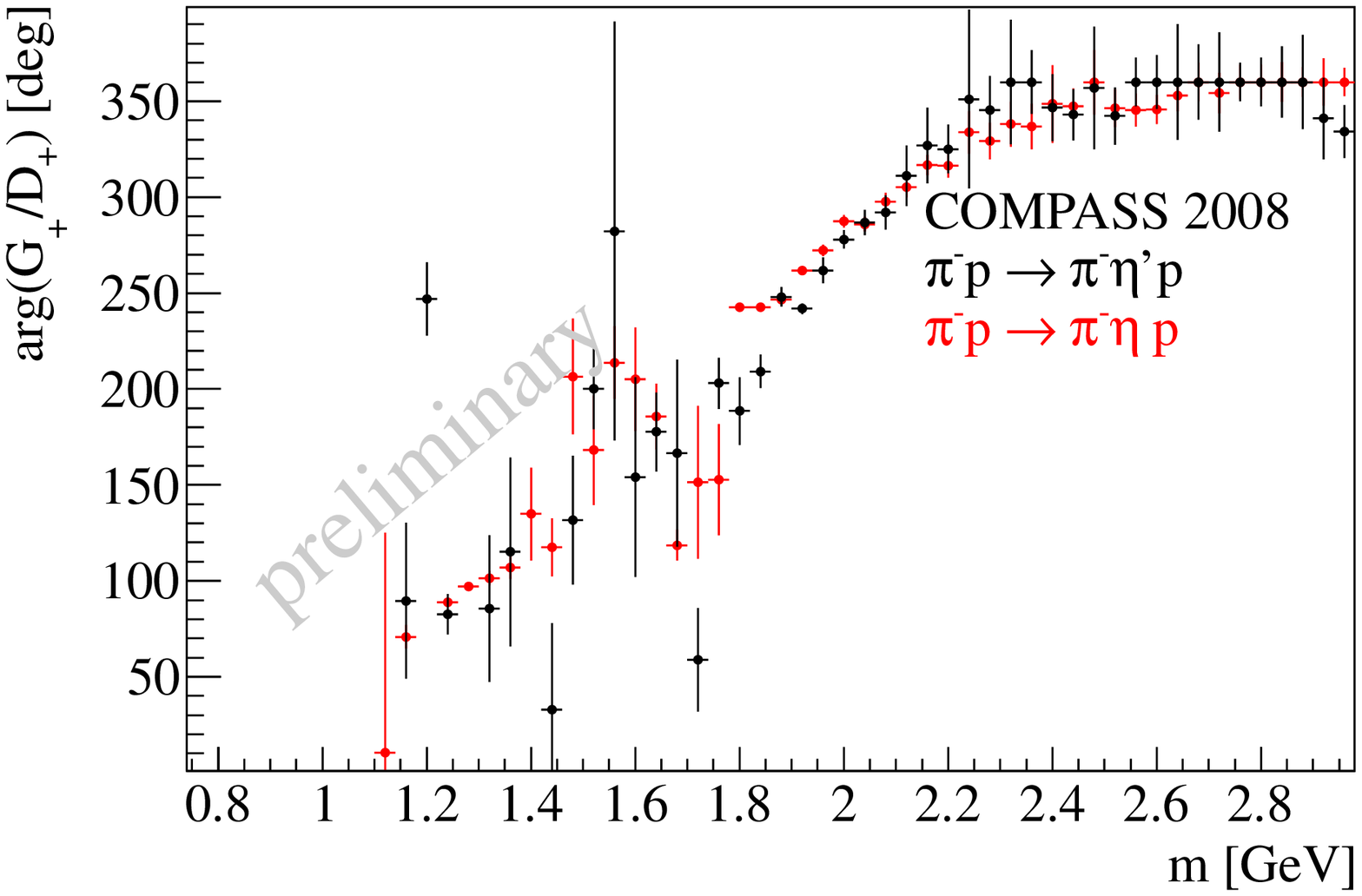}}
  
  \caption[Comparison of relative phases]{Comparison of the relative
    phases.  For the $D_+/P_+$ comparison we show only one of the
    ambiguous branches of the phase-motion in the $\pi^-\eta$ system
    (see text).  The relative phase motion of the $P_+$ and $G_+$
    waves is not shown as they have only very little overlap in the
    $\pi^-\eta$ data.}
  \label{fig:comparison-phases}
\end{figure}

The comparison shows two striking features: first, the close
similarity of the even partial waves, $D_+$ and $G_+$.  The close
match in the overall normalization is supposed to be accidental
subject to further MC studies.  Besides that it appears that the
physical content of these waves is the same in both final states, even
in the high-mass range where non-resonant production is expected to be
dominant.  On the other hand, and the second striking feature, the
$P_+$ wave is strongly suppressed in the $\pi\eta$ final state in
accordance with the suspected non-$q\overline q$ character of this wave and
with a previous analysis by the VES
collaboration~\cite{Beladidze:1993km}.  Comparing the phase motions
(which are not affected by the scaling procedure) as shown in
Fig.~\ref{fig:comparison-phases}, one finds that the $P_+$ wave has
the same phase relative to the $D_+$ wave at the $\eta'\pi$ threshold,
which suggests a common origin, but it then evolves differently which
contradicts them having the same resonant content.  The similarity of
the scaled $D_+$ waves suggests that the difference in the relative
phase motion of the $P_+$ and $D_+$ waves is mainly due to different
contents of the $P_+$ wave.  The aforementioned ambiguity in the phase
determination allows reflecting the extracted phases on the line
corresponding to $-180$ degrees, which would make the relative phase
of the $D_+$ and $P_+$ waves of the $\eta\pi^-$ system return to the
corresponding relative phases of the $\eta'\pi^-$ system at high
masses, suggesting that the difference is due to an incoherent
contribution, which in general tends to reduce relative phase
differences~\cite{Martin:1978jn}.

\section{Outlook and Conclusion}

Beyond what we show here, we have fitted the data with resonance
models.  For the $a_2(1320)$ and $a_4(2040)$ we find parameters that
agree with the PDG~\cite{Nakamura:2010zzi} and other COMPASS
analyses~\cite{Alekseev:2009aa}, respectively.  For a fit to the $P_+$
waves, we need large non-resonant backgrounds to account for both
phase-shifts and intensities simultaneously.  As remarked above, the
phase-shifts seem to indicate that a more complex model allowing for
incoherent contributions is needed.  The studies with higher-spin
waves indicate in particular that non-resonant models should be
explored.  An extraction of the branching fractions of the $a_2(1320)$
and $a_4(2040)$ and comparison to theoretical
predictions~\cite{Bramon:1997va}, while in rough agreement, indicates
that the cross-section of the $\eta'\pi^-$ data is slightly
over-estimated and work is ongoing to understand potential error
sources.

We have performed partial-wave analyses of the $\eta\pi^-$ and
$\eta'\pi^-$ systems.  In these we find as novel results an $m=2$
contribution to the spin-2 wave, we find the $a_4(2040)$ resonance,
and we found a transformation which allows a close comparison of the
even-spin natural-parity partial-wave amplitudes between the two
systems.  A spin-exotic $P_+$-wave contrivution to the two systems
could be confirmed, though its resonant character could not yet be
confirmed unambiguously.

\bibliography{partialwaves,general,statistics,photons}

\end{document}